\documentclass[amssymb,amsmath,prd,twocolumn,preprintnumbers,superscriptaddress,nofootinbib]{revtex4-1}
\usepackage[utf8]{inputenc}
\usepackage{color}
\usepackage{amsmath,bm, physics}
\usepackage{amssymb, mathtools}
\usepackage{hyperref}
\usepackage{cleveref}
\usepackage{graphicx,xcolor}
\usepackage{dcolumn}% Align table columns on decimal point
\usepackage{bm}% bold math
\usepackage{ulem,cancel,soul}

\newcommand{\OmegaGW}{\Omega_{\rm GW}}
\newcommand{\hatOmegaGW}{\hat\Omega_{\rm GW}}
\newcommand{\Omegaref}{\Omega_{\rm ref}}
\newcommand{\zmax}{z_{\rm max}}

\definecolor{mpl_red}{HTML}{D62728}

\newcommand{\pasa}{Publ.\ Astron.\ Soc.\ Pac.}

\begin{document}

\title{Background information: a study on the sensitivity of astrophysical gravitational-wave background searches}

%\title{The sensitivity reach astrophysical gravitational-wave background searches with ground-based detectors}

\author{Arianna I. Renzini}
\email{arenzini@caltech.edu}
\affiliation{LIGO  Laboratory,  California  Institute  of  Technology,  Pasadena,  California  91125,  USA}
\affiliation{Department of Physics, California Institute of Technology, Pasadena, California 91125, USA}
\affiliation{Dipartimento di Fisica “G. Occhialini”, Universit\`a  degli Studi di Milano-Bicocca, Piazza della Scienza 3, 20126 Milano, Italy}
\affiliation{INFN, Sezione di Milano-Bicocca, Piazza della Scienza 3, 20126 Milano, Italy}

\author{Tom Callister}
\email{tcallister@uchicago.edu}
\affiliation{Kavli Institute for Cosmological Physics, The University of Chicago, Chicago, Illinois 60637, USA}

\author{Katerina Chatziioannou}
\email{kchatziioannou@caltech.edu}
\affiliation{LIGO  Laboratory,  California  Institute  of  Technology,  Pasadena,  California  91125,  USA}
\affiliation{Department of Physics, California Institute of Technology, Pasadena, California 91125, USA}

\author{Will M. Farr}
\email{will.farr@stonybrook.edu}
\email{wfarr@flatironinstitute.org}
\affiliation{Department of Physics and Astronomy, Stony Brook University, Stony Brook NY 11794, USA}
\affiliation{Center for Computational Astrophysics, Flatiron Institute, New York NY 10010, USA}

\begin{abstract}
The vast majority of gravitational-wave signals from stellar-mass compact binary mergers are too weak to be individually detected with present-day instruments and instead contribute to a faint, persistent background. %
This astrophysical background is targeted by searches that model the gravitational-wave ensemble collectively with a small set of parameters. %
The traditional search models the background as a stochastic field and estimates its amplitude by \textit{cross-correlating} data from multiple interferometers. %
A different search uses gravitational-wave \textit{templates} to marginalize over all individual event parameters and measure the duty cycle and population properties of binary mergers. %
Both searches ultimately estimate the total merger rate of compact binaries and are expected to yield a detection in the coming years. %
%Given the conceptual and methodological differences between the two searches, the anticipated detection raises questions about interpretation. %
Given the conceptual and methodological differences between them, though, it is not well understood how their results should be mutually interpreted. %
In particular, when a detection of an astrophysical compact binary background is claimed by either approach, which portion of the population is in fact contributing to this detection? %
In this paper, we use the Fisher information to study the implications of a background detection in terms of which region of the Universe each approach probes. %
Specifically, we quantify how information about the compact binary merger rate is accumulated by each search as a function of the event redshift. 
For the LIGO Design sensitivity and a uniform-in-comoving-volume distribution of equal-mass 30$\,M_\odot$ binaries, the traditional cross-correlation search obtains 99\% of its information from binaries up to redshift 2.5 (average signal-to-noise-ratio $<$8), and the template-based search from binaries up to redshift 1.0 (average signal-to-noise-ratio $\sim$8).
While we do not calculate the \textit{total} information accumulated by each search, our analysis emphasizes the need to pair any claimed detection of the stochastic background with an assessment of which binaries contribute to said detection. 
In the process, we also clarify the astrophysical assumptions imposed by each search.
\end{abstract}

\maketitle

%%%%%%%%%%%%%%
\section{Introduction}
%%%%%%%%%%%%%%

%\kc{Things we need to discuss 
%\begin{enumerate}
%\item The $D_{max}$ test in~\cite{Smith:2020lkj}. 
%\end{enumerate}
%}
%\air{on this: it is unclear what prior is used (and up to what redshift) in their production of Fig. 4. From Colm: "I would suspect that they used the same prior as in the rest of the paper, which is likely equivalent to the Bilby UniformComovingVolume". after discussion with Tom, I think we can tone down the comparison between what is done here and in Smith20, saying that their results indicate "some information is contributed by sub-threshold events", but this is not quantified. }

%\todo{Arianna, Tom: the number of citations is very low, are we missing important papers?}

For every compact binary merger~\cite{LIGOScientific:2021djp} observed by the LIGO~\cite{LIGO} and Virgo~\cite{Virgo} observatories, there are many more that are too distant and too weak to be directly detected.
Although these distant binaries may be individually indistinguishable from instrumental noise, their population may be collectively detectable via the slight coherence it imparts across networks of widely separated gravitational-wave detectors~\cite{Allen:1997ad,Romano:2016dpx,KAGRA:2021kbb}.
The resulting collection of weak signals is colloquially referred to as the astrophysical gravitational-wave background. %
Current constraints based on the individually-detectable tail of the total population suggest that this astrophysical background is several orders of magnitude larger than any expected cosmological stochastic background in the relevant frequency range~\cite{KAGRA:2021kbb,PhysRevX.13.011048,Renzini:2022alw}.\footnote{For this reason, we will drop the ``astrophysical" designation in the rest of the paper, with the understanding that unless explicitly noted otherwise we refer to the astrophysical stochastic gravitational-wave background.}
If detected, the stochastic background will offer indirect information about the properties of compact binaries beyond the horizons of present-day detectors~\cite{Callister:2016ewt, Callister:2020arv,Bavera:2021wmw,Lehoucq:2023zlt,Turbang:2023tjk}; it is therefore a prime target for a variety of gravitational-wave searches~\cite{LIGOScientific:2019vic,KAGRA:2021kbb, KAGRA:2021mth}.

Traditional searches for the background (both astrophysical and cosmological) rely on the cross-correlation between detector pairs. %
These searches model the background gravitational-wave strain as a mean-zero stochastic field that is continuous and Gaussian and measure its variance. %
The steep low-frequency noise of ground-based detectors~\cite{aLIGO:2020wna} suggests that any auto-correlations induced by the stochastic background are subdominant and can therefore be neglected; this is often referred to as the weak signal limit~\cite{Allen:1997ad,Romano:2016dpx, Matas:2020roi}. %
But given sufficiently long observation times, the stochastic background will manifest as excess cross power between instruments. %
Given current predictions for the power spectrum of the stochastic background~\cite{KAGRA:2021kbb,PhysRevX.13.011048}, a detection is unlikely in the current observing run, but may be feasible during the next observing run if the Advanced LIGO sensitivity target~\cite{KAGRA:2013rdx} is reached. %

An alternative search strategy is motivated by the fact that the background is neither continuous nor Gaussian. 
Given current merger rate estimates, we expect that a black hole binary (neutron star binary) merges every $5-10$ minutes ($5-60$ seconds) in the mass range relevant for ground-based detectors~\cite{PhysRevX.13.011048}. %
As black hole binary coalescences last for ${\cal{O}}$(seconds) in the ground-based detector frequency band~\cite{LIGOScientific:2018mvr, LIGOScientific:2020ibl, LIGOScientific:2021djp}, the black hole background is expected to be comprised of distinct non-overlapping transient signals. %
For neutron star binaries the duration is ${\cal{O}}$(minutes)~\cite{LIGOScientific:2018hze} so these signals overlap. %
However, given a low frequency cutoff of 10 or 20\,Hz, it is still unlikely that the relevant background lies in the confusion noise limit~\cite{Regimbau:2007cw,Meacher:2014aca,LIGOScientific:2017zlf,Zhong:2022ylh,Johnson:2024foj}, though subject to large uncertainty on the binary neutron star merger rate~\cite{KAGRA:2021kbb}. %
While the non-Gaussianity of the compact binary background does not bias the cross-correlation search in the low-signal limit, it does imply that it is sub-optimal~\cite{Drasco:2002yd}.

Given distinct--though individually undetectable-- transient signals, Ref.~\cite{Smith:2017vfk} proposed a search that relies on the matched-filter technique that has successfully resulted in the detection of individually-resolved binaries, e.g.,~\cite{LIGOScientific:2016vbw}. %
This template-based search utilizes phase-coherent gravitational waveform models to marginalize over the properties of individual events that may (or may not) be present within every segment of data, regardless of whether the events rise above the threshold for direct detection. An extended version also infers and/or marginalizes over the properties of the compact-binary population~\cite{Smith:2020lkj}. %
In general, template-based techniques are expected to be more sensitive than cross-correlation ones as the former include (even vanishingly small) information about the waveform phase, while the latter rely solely on excess power. %
Whereas cross-correlation searches might require years of integration to claim a detection of a gravitational-wave background, it is argued that the template-based search might reach a detection given only \textit{days} of data at design sensitivity~\cite{Smith:2017vfk}. %
If such an improvement is realized, the template-based search would represent a remarkable leap in sensitivity and enable imminent detection.

%The potential gain in sensitivity motivates the main question of this study: 
The stark difference in sensitivity motivates the main question of this study: %
when we detect a background with either search, which region of the Universe have we successfully probed and what astrophysical assumption have we made about this region? %
A direct time-to-detection estimate does not fully address this question as the two searches make different astrophysical assumptions and are not necessarily sensitive to the same population of binaries at the same regions in the Universe. %are impacted by binaries at the edge of LIGO's horizon in different ways. %
Instead, Ref.~\cite{Smith:2020lkj} examined the template-based search's ability to infer a cutoff in the binary redshift distribution and concluded that there is some information from binaries at the edge of resolvability.
Additionally, the cross-correlation and template-based formalisms are constructed in terms of different physical quantities and rely on different detection statistics, specifically the gravitational wave \textit{energy density} and the \textit{event rate} respectively.

The reliance on different statistics is not merely a technical inconvenience, but also reveals a conceptual incompatibility between approaches. %
The energy density and event rate can be mathematically related to each other only after \textit{assuming a specific merger rate distribution} for sources across the Universe. Gravitational-wave detectors measure spacetime strain, which can be trivially converted to intensity and energy density without further assumptions. This suggests that the cross-correlation search can infer the gravitational-wave energy density directly from the data, with no assumption on source distribution.  
The template-based analysis, on the other hand, is based on the measurement of a non-zero rate of events. The conversion from measured strain to an event rate (or duty cycle) relies on a Bayes factor that compares the hypotheses that a data segment contains signal or noise, which in turn depends on assumptions about the prior or astrophysical distribution of sources, including the mass, spin, and redshift distribution. While Ref.~\cite{Smith:2020lkj} relaxes the mass and spin dependence, the assumption of knowledge of the redshift distribution remains. 
If an analysis intrinsically assumes perfect knowledge of how sources are distributed in the Universe, this begs the question: how much information is actually coming from the unresolved sources compared to what is extrapolated from the resolved ones?

The goal of our work is to study the sensitivity of the cross-correlation and template-based searches and identify the astrophysical assumptions each search makes. 
We use the Fisher Information Matrix to quantify the information accumulated by each search and study how this information is accumulated by observing binaries at different redshifts. 
In lieu of a technically challenging full implementation of the template-based search, we consider a simplified population of compact binaries where all parameters are known other than the distance/redshift.
This simplified set up does not allow us to compare the \textit{total} information accumulated by each search, but it yields the \textit{relative} contribution by binaries at different redshifts.
For $30\,M_{\odot}$ binaries, the cross-correlation (template-based) search accumulates 99\% of its information from binaries up to redshift $2.5\,(1.2)$. The same redshift estimates for $70\,M_{\odot}$ binaries are 1.5 (1.6).
With the hopefully imminent detection of a stochastic background in LIGO data, our study emphasizes the need to carefully assess the sensitivity and assumptions of our search methods.

The rest of the paper presents the details of our calculation.
We begin in Sec.~\ref{sec:background} by reviewing the stochastic background and its characterization.
In Sec.~\ref{sec: methods} we lay out the two search methodologies, highlighting the targeted observables. In Sec.~\ref{sec:information} we describe the theory and calculate the functional forms of the Fisher information for each search with respect to a common parameter, and in Sec.~\ref{sec:I-study} we study the information functions in different scenarios. Finally, we draw our conclusions in Sec.~\ref{sec:conclusions}.

%%%%%%%%%%%%%%%%%%%%%%%%%%%%%%%%%%%%%%%%%%%%%%%
\section{The compact binary background}
\label{sec:background}
%%%%%%%%%%%%%%%%%%%%%%%%%%%%%%%%%%%%%%%%%%%%%%%

In this section, we introduce the basics of the gravitational-wave stochastic background and establish notation. The familiar reader can skip ahead to Sec.~\ref{sec: methods} where we introduce the search methods.

The background of unresolved compact binaries is generally characterized by its dimensionless energy-density spectrum~\cite{Allen:1997ad,Renzini:2022alw},\footnote{This formalism is largely inspired by the cosmological background, but it is applied to the astrophysical background as well.}
\begin{equation}
	\label{eq:Omega}
	\OmegaGW(f) = \frac{1}{\rho_c}\frac{d\rho_\textsc{gw}}{d\ln f}\,,
\end{equation}
where $\rho_\textsc{gw}$ is the energy-density in gravitational waves and $\rho_c$ is the Universe's closure energy density. %
The present-day $\OmegaGW(f)$ in the frequency range $10\,\mathrm{Hz}<f<10^3$\,Hz is dominated by the integrated merger history of compact binaries over all redshifts. %
For simplicity, we consider a single class of compact binaries with ensemble-averaged source-frame energy spectra $dE_s/df_s$ (a subscript $s$ denotes source-frame quantities), and define ${\cal R}(z)$ to be the number of mergers per unit comoving volume $V_c$ per unit source-frame time $t_s$, i.e., the source-frame merger rate density
\begin{equation}
    {\cal R}(z) = \frac{dN}{dV_c dt_s}\,.
\end{equation}
Given gravitational-wave sources up to redshift $z_\mathrm{max}$, the present-day dimensionless energy-density spectrum is~\cite{KAGRA:2021kbb}
\begin{equation}
\label{eq:stoch-energy-density}
\begin{aligned}
\OmegaGW(f)
    &= \frac{f}{\rho_c} \int_0^{z_\mathrm{max}} {\cal R}(z) \frac{dE_s}{df_s}\Big|_{f(1+z)} \frac{dt_s}{dz} dz \\
    &= \frac{f}{\rho_c} \int_0^{z_\mathrm{max}} \frac{{\cal R}(z) }{(1+z) H(z)} \frac{dE_s}{df_s}\Big|_{f(1+z)} dz\,,
\end{aligned}
\end{equation}
where $H(z) \approx H_0\sqrt{\Omega_M (1+z)^3 + \Omega_\Lambda}$ is the Hubble parameter at redshift $z$ and $H_0$ is the Hubble constant. %
%%%%
The total detector-frame merger rate $R$ is
\begin{equation}
	R = \int_0^{z_\mathrm{max}} dz\,\frac{\mathcal{R}(z)}{1+z} \frac{dV_c}{dz}\,.
    \label{eq:R_tot_integral} 
\end{equation}
Here $dV_c/dz$ is the comoving volume per unit redshift.
The factor of $1+z$ converts the source-frame rate into the detector-frame rate as measured at Earth.
%%%%

Both the gravitational-wave energy-density and the total merger rate depend on the source-frame rate density, $\mathcal{R}(z)$.
It is convenient to write the latter as the product $\mathcal{R}(z)=R_0 r(z)$, where $R_0$ is the local merger rate at $z=0$ and $r(z)$ is the merger rate density function normalised to 1 at $z=0$.
Then, in terms of the total rate $R$, the local merger rate becomes
\begin{equation}
	\label{eq:R0-helper}
    \begin{aligned}
	R_0 &= R \left[\int_0^{z_\mathrm{max}} dz\,\frac{r(z)}{1+z} \frac{dV_c}{dz}\right]^{-1} \\
    &\equiv R \,{\cal I}(\zmax)^{-1}\,,
    \end{aligned}
\end{equation}
where we have defined the integral ${\cal I}(\zmax)$ which quantifies the ratio between total and local merger rate, given $\zmax$. We also define the normalized probability distribution for source redshift, $p(z)$,
\begin{equation}
    \begin{aligned}
    p(z) &= \frac{\frac{\mathcal{R}(z)}{1+z} \frac{dV_c}{dz}}{\int_0^{z_\mathrm{max}} dz' \frac{\mathcal{R}(z')}{1+z'} \frac{dV_c}{dz'}} \\
    &= \frac{R_0}{R} \frac{r(z)}{1+z} \frac{dV_c}{dz} \\
    &\equiv \frac{1}{R} R(z)\,,
    \end{aligned}
    \label{eq:p_of_z}
\end{equation}
defining $R(z)$ to be the event rate per unit detector-frame time.

So far, searches targeting the stochastic background via $\OmegaGW(f)$ have operated under the assumption that the signal is continuous and Gaussian. However,
the background is only Gaussian in the limit of large numbers of sources that saturate the detector time-stream and satisfy the central limit theorem. Given the current set of compact binary detections~\cite{PhysRevX.13.011048}, it is clear that this is not the case~\cite{LIGOScientific:2017zlf}. %
As merging black holes are expected to strongly contribute to the GW energy density~\cite{Regimbau:2011rp}, this suggests the astrophysical stochastic background is non-Gaussian and intermittent in the detector frequency band, with a cadence dictated by the prevalence of black hole binary mergers in the Universe. %

To quantify the non-Gaussianity and intermittence of the signal and inspired by~\cite{Drasco:2002yd}, Ref.~\cite{Smith:2017vfk} split the data in segments of duration $\tau$, and denoted the probability that a given segment contains a signal as $\xi$. %
This duty cycle $\xi$ can be related to the merger rate. Given the total detector-frame merger rate $R$, the probability that $N_{\rm s}$ gravitational-wave signals are present in a data segment with duration $\tau$ is Poisson-distributed
\begin{equation}
    p(N_{\rm s} | R,\tau) = \frac{\left(R\tau\right)^{N_{\rm s}} e^{-R\tau}}{N_{\rm s}!}\,.
\end{equation}
Then the duty cycle $\xi$ is simply the probability that $N\geq1$ signals are present in the data segment:
\begin{equation}
    \label{eq:xi-helper}
    \begin{aligned}
\xi &= \sum_{N_{\rm s}=1}^\infty p(N_{\rm s} | R,\tau) \\
    &= 1 - p(0|R,\tau) \\
    &= 1 - e^{-R_0 {\cal I}(\zmax)\tau}\,.
    \end{aligned}
\end{equation}
In the limit where $R\tau \ll 1$, this becomes simply $\xi~\approx~R\tau$. %
Assuming a segment length $\tau\sim{\cal{O}}(\mathrm{s})$\footnote{With current detector sensitivity, and in order to avoid double-counting, it is reasonable to pick a segment duration such that there is at most one event per data segment. The segment duration is therefore chosen to be comparable to the time a signal spends in the detector frequency band.}, the expected duty cycle of the black hole background is $\xi\sim 10^{-3}$.

The cross-correlation analysis (described in Sec.~\ref{sec: methodsCC}) is typically expressed in terms of the $\OmegaGW(f)$ spectrum, as it was conceived for Gaussian and continuous backgrounds. %, i.e., the average energy-density received by the detectors is constant and $\langle \OmegaGW \rangle_t = \langle{\Omega}_{{\rm GW}}\rangle_p$. 
The template-based search (described in Sec.~\ref{sec: methodsTB}), on the other hand, was proposed with an non-Gaussian and intermittent background in mind and is hence framed in terms of $\xi$. %
In Sec.~\ref{sec:information} 
we also work in terms of independent parameters which the observables of both searches can be mapped to. Specifically, we use Eqs.~\eqref{eq:stoch-energy-density},~\eqref{eq:R_tot_integral} and~\eqref{eq:xi-helper} to express the cross-correlation and template-based searches in terms of the same quantities, $R_0$ and $\zmax$, for direct comparison. %

%%%%%%%%%%%%%%
\section{Stochastic background search methods}
\label{sec: methods}
%%%%%%%%%%%%%%

We consider two searches that target the sub-threshold population of binary mergers in ground-based detectors: the cross-correlation search (relevant quantities are labelled as ``CC") and the template-based search (``TB").

%------------------------------------------------------------------------------------
\subsection{Cross-Correlation Search}
\label{sec: methodsCC}

The most common search for the stochastic gravitational-wave background relies on the cross-correlation spectrum between the (frequency domain) data $\tilde d_1(f)$ and $\tilde d_2(f)$ measured by two gravitational-wave detectors to construct an optimal (i.e., unbiased minimum-variance) statistic for $\OmegaGW(f)$~\cite{Allen:1997ad,Romano:2016dpx,Matas:2020roi},
\begin{equation}
\label{eq:cross-corr}
	\hat{\Omega}_{\rm GW}(f) = \frac{2 Q(f)}{\gamma(f)} \frac{\mathbb{Re}\qty(\tilde d_1(f) \tilde d_2^*(f))}{T_{\rm seg}}\,,
\end{equation}
where $T_{\rm seg}$ is the time segment duration over which data are measured. %
Here and throughout this discussion we use ``hat-less'' symbols to denote physical quantities and ``hatted'' symbols to denote their estimate based on data.
The data are Fourier transforms of strain data and hence have units $[\,\tilde{d}\,] = {\rm Hz}^{-1}$. The function $Q(f)$ is defined as
\begin{equation}
    Q(f) = {f^3}  \frac{10 \pi^2}{3 H_0^2}\,,
\end{equation} 
and converts the strain power spectrum to a dimensionless energy-density spectrum.
The factor $\gamma(f)$ is the overlap reduction function~\cite{1992PhRvD..46.5250C,flanagan_sensitivity_1993}, which quantifies the geometrical sensitivity of the cross-correlated detector pair to an isotropic background.
The factor of $2$ in Eq.~\eqref{eq:cross-corr} accounts for the contribution of negative frequencies.
The variance of the statistic in Eq.~\eqref{eq:cross-corr} is~\cite{KAGRA:2021kbb}
\begin{equation}
	\hat{\sigma}^2_{\rm GW}(f) = \frac{1}{2} \qty(\frac{Q(f)}{\gamma(f)})^2 \hat{P}_1(f) \hat{P}_2(f)\,,
\end{equation}
where $\hat{P}_1(f)$ and $\hat{P}_2(f)$ are the one-sided strain power spectra of the data in detectors 1 and 2, respectively, defined as
\begin{equation}
    \hat{P}_I(f) = \frac{2}{T_{\rm seg}} \abs{\tilde{d}^{}_I(f)}^2\,. %\equiv \Delta f^{-1}\,  {\cal P}_{i, \rm tot}
    \label{eq:PIs}
\end{equation}

The energy density spectrum can be decomposed as
\begin{equation}\label{eq:Omega_ref}
    \OmegaGW(f) = \Omegaref\, E(f/f_{\rm ref})\,,
\end{equation}
 with an overall amplitude $\OmegaGW(f_{\rm ref}) \equiv \Omegaref$ at reference frequency $f_{\rm ref}$. The spectral shape $E(f/f_{\rm ref})$ can be assumed known or parametrized $E(f/f_\mathrm{ref})$ and inferred~\cite{Mandic:2012pj}. For example, for compact binary sources $E(f/f_\mathrm{ref})$ should be universal, as the inspiral frequency evolution is independent of $r(z)$. This is true up to a turn-over frequency which corresponds to the redshifted merger frequency of the binaries. %
 If we are not sensitive to the spectrum turn-over, as is the case with current ground-based interferometers, we can treat the spectral shape of the signal as redshift-independent and set $E(f/f_\mathrm{ref}) \propto (f/f_\mathrm{ref})^{2/3}$.

In practice, cross-correlation spectra are estimated independently for a large number of short time segments, each of ${\cal{O}}(100\,\mathrm{s})$, and combined via a weighted average. %
Since a large number ($10^4-10^5$) of such time segments are combined, the resulting cross-correlation measurements $\hatOmegaGW(f)$ are well-described by Gaussian statistics even in the case of an intermittent signal. %
As shown in~\cite{Matas:2020roi}, this cross-correlation statistic is a sufficient statistic for the stochastic signal in the case where the signal is Gaussian and weak compared to detector noise\footnote{A weak signal is a signal that does not appreciably contribute to the power measured by a single detector~\cite{Romano:2016dpx}, such that the variance of the data can be equated to the variance of the noise.}, and can be derived from the mean-zero Gaussian likelihood
%\todo{Arianna, Tom says ``I think we should remain general (i.e only showing things in terms of $\Omega(f))$ until after Eq. 18, where we can then say that often we further simplify things by treating $\Omega(f)$ as some amplitude parameter times a constant shape function." Again your call}
%
\begin{equation}
    p^{\rm CC}( \tilde{d}|\OmegaGW, P_{n,I})  = \prod_f \frac{1}{|2\pi{\bm C}(f)|} e^{-\frac{1}{2} \tilde{d}^\dagger {\bm C}^{-1} \tilde{d}}\,.
    \label{eq:cross-corr-full-like}
\end{equation}
Here, $\tilde{d}$ is the data vector $\tilde{d} = \,
(\tilde{d}_1, \,\tilde{d}_2)^T$, and $\OmegaGW(f)$ appears in the data covariance as the intrinsic variance of the gravitational-wave signal~\cite{Matas:2020roi},
\begin{equation}
    {\bm C}(f) = \frac{T_{\rm seg}}{4}\begin{pmatrix}
    P_1(f) & \gamma(f)P_{\rm GW}(f) \\
    \gamma(f)P_{\rm GW}(f)  & P_2(f)
    \end{pmatrix}\,,
    \label{eq:covariance}
\end{equation}
where 
\begin{equation}
    \label{eq:PGW}
    P_{\rm GW}(f) =  \frac{\OmegaGW(f)}{Q(f)}\equiv \Omegaref\frac{E(f/f_{\rm ref})}{Q(f)}\,.
\end{equation}
For stationary noise within each analysis segment the noise covariance is diagonal in frequency, hence the total likelihood is a product of Gaussian likelihoods evaluated at each frequency. %
The $P_I(f)$ parameters are the expected power spectra in each detector,
and can be explicitly written as the sum of the noise and gravitational-wave power spectra, 
\begin{equation}
    P_I(f) = P_{n,I}(f) + P_{\rm GW}(f)\,.
    \label{eq:PSD-data}
\end{equation}
This implies both diagonal and off-diagonal terms of the covariance in Eq.~\eqref{eq:covariance} depend on the gravitational-wave energy density, giving rise to autocorrelation and cross-correlation terms respectively. The noise power spectra $P_{n,I}(f)$ are additional parameters of the search, and may in principle be estimated alongside $\OmegaGW(f)$.

In the weak signal limit, the auto-correlation terms in Eq.~\eqref{eq:cross-corr-full-like} can be neglected, such that $P_I(f) \approx P_{n,I}(f)$ and the likelihood reduces to %
\begin{equation}
    \label{eq:singleCClikelihood}
    p^{\rm CC}(\hatOmegaGW | \OmegaGW ) \propto \prod_f\exp\qty[-\frac{\left(\hatOmegaGW(f) - \OmegaGW(f)\right)^2}{2\hat{\sigma}_{\rm GW}^2(f)}]\,,
\end{equation}
where hatted quantities are directly estimated from the data. %
To simplify parameter estimation, we fix $E(f/f_{\rm ref}) = (f/f_{\rm ref})^{2/3}$ such that 
the likelihood of Eq.~\eqref{eq:singleCClikelihood} refers to a single parameter, $\Omegaref$, and depends only on quantities derived from the data as well as the assumed $E(f)$. %, which is universal for black hole binaries. %
It is possible to remove this second dependence by re-expressing the likelihood in the full spectrum $\OmegaGW (f)$, and constrain each frequency bin independently, however $\Omegaref$ is typically preferred as this allows us to marginalize over the spectrum and improve detection statistics.

%------------------------------------------------------------------------------------
\subsection{Template-based search}
\label{sec: methodsTB}

The template-based search adopts a different approach that is more similar to the traditional matched-filter searches for individually-detectable signals~\cite{Smith:2017vfk,HernandezVivanco:2019fku,Smith:2020lkj}. 
In this search, the entire strain time series measured by a gravitational-wave detector network is divided into $N_t$ time segments, each with duration $\tau$. %
For instance, the search for a stochastic background from black-hole binaries divides one year of data into approximately $N_t=10^7$ segments of $\tau = 4$\,s duration, given the reasonable expectation that each segment contains~$\ll 1$ signal on average~\cite{Smith:2017vfk}. %
Here the segment length is selected by considering the expected single-event duration at present sensitivity. %
This is in stark contrast with the typical choices made for the cross-correlation analysis, where the segment duration is typically on the time scale of a few minutes to access lower frequency content, handle noise non-stationarity, and minimize the computational cost of the search~\cite{Renzini:2023qtj}.

Within every time segment $i$, a template-based analysis computes the marginalized likelihood (or ``evidence'') 
that a compact binary merger is (hypothesis $\mathcal{S}_i$) or is not (hypothesis $\mathcal{N}_i$) present. %
The marginalized likelihood for the signal hypothesis is obtained via marginalization over all source parameters $\theta$ of the binary,
\begin{equation}
    p^{\rm TB}(d_i | \mathcal{S}_i) = \int p(d_i |\theta,\mathcal{S}_i)\, p(\theta | \mathcal{S}_i) d\theta\,,
\end{equation}
where $d_i$ is the data comprising segment $i$, $p(d_i | \theta,\mathcal{S}_i)$ is the likelihood of having obtained this data in the presence of a source with parameters $\theta$ (binary masses, spins, redshift, etc.), and $p(\theta | \mathcal{S}_i)$ is the prior on the source parameters $\theta$. %

With the segment-by-segment signal and noise evidences in hand, the template-based analysis then seeks to measure the duty cycle $\xi$, i.e., the fraction of time segments containing a gravitational-wave signal, see discussion in Sec.~\ref{sec:background}.
The relevant likelihood across all $N_t$ segments is
\begin{align}
	\label{eq:xi-likelihood-A}
	p^{\rm TB}(\{d_i\}|\xi)
		&= \prod_i p(d_i|\xi) \nonumber \\
		&= \prod_i \Bigl[ p(d_i | \mathcal{S}_i) p(\mathcal{S}_i | \xi) + p(d_i | \mathcal{N}_i) p(\mathcal{N}_i | \xi) \Bigr] \nonumber\\
		&= \prod_i \Bigl[ \xi \,p(d_i | \mathcal{S}_i) + (1-\xi) \,p(d_i | \mathcal{N}_i) \Bigr]\,,
\end{align}
where, by definition, $p(\mathcal{S}_i | \xi) = \xi$ is the probability that a signal is present in segment $i$; correspondingly $p(\mathcal{N}_i | \xi) = 1-\xi$ is the probability that a signal is absent.
Factoring out $p(d_i | \mathcal{N}_i)$, we can rewrite the likelihood as
\begin{equation}
    \label{eq:xi-likelihood-B}
    p^{\rm TB}(\{d_i\}|\xi) = \prod_i p(d_i | \mathcal{N}_i) \left[ \xi b_i  + (1-\xi) \right]\,,
\end{equation}
where $b_i$ is the Bayes factor between signal and noise hypotheses in segment $i$:
	\begin{equation}
	b_i = \frac{p(d_i | \mathcal{S}_i)}{p(d_i | \mathcal{N}_i)}\,.
	\end{equation}
Within the framework of the template-based search, observation of the stochastic background amounts to constraining $\xi$ away from zero.
Equivalently, the Bayes factor $\mathcal{B}_\xi$ between the ``signal'' hypothesis that allows $0\leq\xi\leq1$ and the ``noise'' hypotheses in which  $\xi=0$
	\begin{equation}
	\label{eq:tbs-bayes}
	\mathcal{B}_\xi = \frac{\int_0^1 p(\{d_i\} | \xi) p(\xi) d\xi}{p(\{d_i\}|\xi=0)}\,,
	\end{equation}
can be used as a detection statistic given some prior on the duty cycle $p(\xi)$. 

In contrast to Eq.~\eqref{eq:singleCClikelihood}, which only depends on the data and the inferred quantities, Eq.~\eqref{eq:tbs-bayes} also depends on a collection of binary parameter priors. These include priors on compact binary masses $m$, spins $\chi$, and redshift $z$ (or, equivalently, distance). 
These priors encode our belief about the underlyling population of compact binaries.
Choosing a particular prior $p(m,\chi,z | \mathcal{S}_i)$ when evaluating Eq.~\eqref{eq:xi-likelihood-B} amounts to assuming that the population distribution of these parameters is perfectly known. %
An extended version of the template-based analysis relaxes the assumption of a known population distribution for masses and spins~\cite{Smith:2020lkj}.
Instead, the population prior $p(m,\chi | \mathcal{S}_i)$ is parametrized and the resulting set of hyperparameters are added to the search. %
Conceptually, this is equivalent to performing a search for compact binary mergers while simultaneously performing a population analysis to measure their ensemble properties~\cite{LIGOScientific:2020kqk,PhysRevX.13.011048}. % 
However a fixed prior on binary redshift remains: the current formulation of the search assumes perfect knowledge of how compact binaries are distributed with redshift. %
This assumption, combined with the presence of resolved low-redshift binaries, raises the question of whether a measurement of $\xi>0$ is informed by binaries at high redshift, or is instead dominated primarily by foreground binaries.

%raises the question of how much these foreground binaries dominate the search and how much information is actually received from higher redshift. %
This question motivates our study: below we revisit the sensitivity of stochastic searches and quantify the impact of resolved sources in the template-based case.  %

%%%%%%%%%%%%%%%
\section{Information content}
\label{sec:information}
%%%%%%%%%%%%%%%

The different assumptions, methodology, and formalism between the cross-correlation and template-based searches make a direct comparison in terms of a simple time-to-detection difficult. 
We instead consider the \textit{Fisher information} for each search and examine how this information is gathered as a function of source redshift and for different astrophysical assumptions about the event distribution.
Given a likelihood $p(d|\Lambda)$ for data $d$ conditioned on parameters $\Lambda=\{\Lambda_i\}$, the Fisher information matrix is defined as the expectation value over data realizations
	\begin{equation}
 \label{eq:Fisher-matrix}
	F_{ij}(\Lambda) = \left\langle -\frac{\partial^2}{\partial \Lambda_i\partial \Lambda_j} \ln p(d|\Lambda) \right\rangle\,.
	\end{equation}
In the high signal-to-noise ratio limit where the likelihood becomes approximately Gaussian, the inverse Fisher matrix $F^{-1}_{ij}$ corresponds to the covariance matrix quantifying the uncertainties on parameters $\Lambda_i$.
The Fisher information, meanwhile, is defined as the matrix determinant:
\begin{equation}
    I(\Lambda) = {\rm det}\,F_{ij}(\Lambda)\,.
\end{equation}
The strong signal-to-noise assumption of the Fisher formalism is not in tension with the previously-defined low-signal limit employed for the cross-correlation search. 
The latter refers to the fact that the auto-correlation of strain data is dominated by detector noise, rather than astrophysical signals. The former, in contrast, refers to the detectability of excess cross-power due to the gravitational-wave background, after integrating over a sufficiently long period of time. %
In other words, even though the individual signals are sub-threshold, the applicability of the Fisher formalism refers to the stochastic signal as a whole.

The cross-correlation and template-based searches are framed in terms of different observables, i.e., $\Lambda$ is different in each case.
The cross-correlation search reports a measurement of the amplitude $\Omegaref$, i.e., $\Lambda^{\rm CC}=\Omegaref$. %
%Detection, then, constitutes the exclusion of $\Omegaref = 0$. %
The template-based search, meanwhile, reports a measurement of the duty cycle $\xi$, i.e., $\Lambda^{\rm TB}=\xi$.
%; here detection amounts to excluding $\xi=0$. %
The duty cycle of a population and its total fractional energy density can be related by combining Eqs.~\eqref{eq:stoch-energy-density},~\eqref{eq:R0-helper}, and~\eqref{eq:xi-helper}, however this requires prior knowledge of the source redshift distribution $r(z)$. %
Equivalently, knowledge of $r(z)$ is required to convert the fractional energy density emitted by a population to its local merger rate density $R_0$, and therefore to $\xi$. %
Comparing the two searches thus relies on prior knowledge of $r(z)$, which raises doubts as to what is actually being measured, as opposed to extrapolated, to produce the individual search results as well as in their comparison. %
To make progress, our strategy here is to parametrize $r(z)$ and infer it alongside each search parameters. %

Extending each search to also probe the source redshift distribution amounts to a parameterization of $r(z) \equiv r(z;\lambda)$  and additional parameters $\lambda$. %
Here, we adopt the redshift distribution parametrization of Eq.~\eqref{eq:R0-helper}, with $\lambda=\zmax$, similarly to~\cite{Smith:2020lkj}. %
For the template-based search this results in $\Lambda^{\rm TB}=\{\xi,\zmax\}$. %
In terms of the likelihood in Eq.~\eqref{eq:xi-likelihood-B}, the duty cycle $\xi$ is explicit, while $\zmax$ enters through the distance/redshift prior used in the Bayes factor calculation, $b_i(\zmax)$.
However, varying $\zmax$ while keeping $\xi$ constant would result in a change in the redshift distribution of the local, individually-detectable sources. Since this is expected to be well-constrained by the time of detection of the stochastic background, we instead reparametrize the analysis to explicitly separate the local merger rate $R_0$ from the source redshift distribution, $\Lambda^{\rm TB}=\{R_0,\zmax\}$. %

In the case of the cross-correlation search, the only parameter considered in the likelihood of Eq.~\eqref{eq:singleCClikelihood} is $\Omegaref$.
For direct comparison with the template-based search, we also need to express this likelihood in terms of the same $R_0$ and $\zmax$ parameters, linked to $\Omegaref$ through Eqs.~\eqref{eq:stoch-energy-density},~\eqref{eq:R0-helper}. We therefore also consider a $2d$ version of the cross-correlation search with $\Lambda^{\rm CC}=\{R_0,\zmax\}$, though both parameters enter the likelihood only through their $\Omegaref$ combination.
In what follows, we discuss both the case where the maximum cutoff $\zmax$ is assumed known and fixed ($1d$ searches), and the case where it is a free parameter ($2d$ searches). %
For the $1d$ comparison, we present results in $\Omegaref$, as this naturally quantifies the background amplitude; while for the $2d$ case, we use $R_0$ and $\zmax$ for the reasons detailed above. %

A back-of-the-envelope calculation clarifies the ensuing detailed calculation in the $1d$ case.
Here, the cross-correlation search is sensitive to the GW energy density while the template-based one is sensitive to the local merger rate. 
In the cross-correlation case, the energy density of $N$ events contained in a volume $\propto D^3$ scales as $ N \times E \sim D$, as the single event energy is $E\propto D^{-2}$. This means that the total energy density scales roughly linearly with event distance. Meanwhile, for the template-based search the  likelihood is dominated by the $\xi\times b$ term, which scales as $\xi\times b \propto N\times e^{{\rm SNR}^2/2} \sim D^{3} \times e^{D^{-2}/2}$, where $b$ is an event Bayes factor as in~\eqref{eq:xi-likelihood-B}, and we have assumed that $b\propto e^{{\rm SNR}^2/2}$ implying a loud signal. This indicates that events at large $D$ will contribute less to this search, as the Bayes factor decreases rapidly with distance. In what follows, to avoid taking this loud signal approximation for events around/below detection threshold, we evaluate Bayes factors numerically. 

%%%%%%%%%%%%%%%%%%%%%%%%%%%%%%%%%%%%%%%%%%
\subsection{Cross-Correlation Search}

We calculate the Fisher information for the cross-correlation search starting from the full likelihood of Eq.~\eqref{eq:cross-corr-full-like} which may be expanded as~\cite{Matas:2020roi}
\begin{widetext}
\begin{equation}
    \ln p^{\rm CC}(\tilde{d}|\Omegaref) \sim - N_{\rm seg} \sum_f \qty( \ln |{\bm C}(f)| + \qty(\frac{T_{\rm seg}}{4})^2 \frac{P_1(f)\hat{P}_{2}(f) + P_2(f)\hat{P}_{1}(f) - 2 {\gamma^2(f)}{Q^{-2}(f)}\Omegaref E(f/f_\mathrm{f}) \hat{\Omega}_{\rm GW}(f)}{|{\bm C}(f)|})\,,
\end{equation}
\end{widetext}
where the determinant of the covariance is
\begin{equation}
    |{\bm C}(f)| = \qty(\frac{T_{\rm seg}}{4})^2 \qty[P_1(f) P_2(f) - \gamma^2(f)  P_{\rm GW}^2(f)] \,,
\label{eq:detC}
\end{equation}
and $N_{\rm seg}$ is the total number of segments used in the analysis. %
In what follows, we consider the noise terms $P_{n,I}$ entering the data power spectra $P_I$ in Eq.~\eqref{eq:PSD-data} to be known.

 We proceed to calculate the Fisher matrix in $\log\Omegaref$ and $\Omegaref$, $F_{\log\Omegaref}$ and $F_{\Omegaref}$ respectively. %
 Here $F_{\log \Omegaref}$ corresponds to the fractional uncertainty on $\Omegaref$, and thus scales with the sensitivity of the search. In contrast, $F_{\Omegaref}$ scales with both the sensitivity and the value of $\Omegaref$ itself. %
 In terms of $\log\Omegaref$ we have
\begin{align}
	F^{\rm CC}_{\log\Omegaref} &= \left\langle - \frac{\partial^2}{\partial \log\Omegaref^2} \ln p^{\rm CC}(\tilde{d}|\Omegaref) \right\rangle\,,
    \label{eq:ICC_logOm}
\end{align}
where the angle brackets imply calculating the expectation value at maximum likelihood over many realizations, hence first derivatives of the likelihood are dropped. %

Under the noise stationarity assumption, each frequency contributes to the likelihood calculation independently, hence the total contribution is the sum over individual frequencies:
\begin{equation}\label{eq:I_logOmega}
    F^{\rm CC}_{\log \Omegaref}=\Omegaref^2 \sum_f E^2(f)F^{\rm CC}_{\Omega_{f}}\,,
\end{equation}
where 
\begin{equation}
    F^{\rm CC}_{\Omega_{f}} = \left\langle - \frac{\partial^2}{\partial \Omega_f^2} \ln p^{\rm CC}(\tilde{d}|\Omega_f) \right\rangle\,,
\end{equation}
computed at the individual frequency $f$, where $\Omega_f = \Omegaref E(f)$. Taking the maximum likelihood limit (which amounts to setting hatted quantities equal to their unhatted counterparts, assuming these are unbiased estimators), defining $\beta(f) = \gamma(f)^2 - 1$, and dropping the explicit frequency dependence from the functions $\gamma$, $\beta$, $P_{n,I}$, $E$, and $Q$ for conciseness, we find
\begin{widetext}
\begin{equation}
    F^{\rm CC}_{\Omega_f} = N_{\rm seg} E^2 \frac{2 \beta^2 \Omega_f^2 -2 \beta Q (P_{n,1} + P_{n,2}) \Omega_f +Q^2 \qty(P_{n,1}^2+2
   \gamma ^2 P_{n,1} P_{n,2}+P_{n,2}^2) }
   {\bigl(P_{n,1} Q ( \Omega_f +Q P_{n,2})+ \Omega_f  \qty(Q P_{n,2}-\beta \Omega_f) \bigr)^2}\,.
\end{equation}
\end{widetext}
In accordance with the standard cross-correlation search, we take the low-signal limit and expand $F^{\rm CC}_{\Omega_{f}}$ in $\epsilon = \gamma \Omega_f/(F\sqrt{P_{n,1} P_{n,2}})$, which yields to first order,
\begin{equation}
    F^{\rm CC}_{\Omega_{f}} = A(f) + B(f)\,\Omega_f + {\cal{O}}(\epsilon^2)\,,
    \label{eq:stoch-total-info-absolute}
\end{equation}
where
\begin{align}\label{eq:Af}
    A(f) &= N_{\rm seg} \frac{
   \qty(P_{n,1}^2+2 \gamma ^2 P_{n,1}P_{n,2}+P_{n,2}^2)}{ Q^2
   P_{n,1}^2 P_{n,2}^2}\,,\\
   B(f) &= - 2N_{\rm seg} \frac{\qty(P_{n,1}^3+3 \gamma ^2 P_{n,1}^2 P_{n,2}+3 \gamma ^2
   P_{n,1}P_{n,2}^2+P_{n,2}^3)}{ Q^{3} P_{n,1}^3 P_{n,2}^3}\,.\label{eq:Bf}
\end{align}
The term $F^{\rm CC}_{\log \Omegaref}$ is then
\begin{equation}\label{eq:I_logOmegaref}
    F^{\rm CC}_{\log \Omegaref}=\Omegaref^2 \sum_f E^2(f)\qty[ A(f) + B(f)\,E(f) \, \Omegaref ]\,.
\end{equation}
Since we are restricting to a single parameter here, Eq.~\eqref{eq:I_logOmegaref} gives the Fisher information gathered in a cross-correlation search for the fractional uncertainty in $\Omegaref$, 
and has two limiting cases. %
In the limit where the signal $\Omegaref\rightarrow 0$, i.e., a Universe with no compact binaries, the absolute information per frequency bin is $F^{\rm CC}_{\Omega_f} \rightarrow A(f)$. Hence, $A(f)$ 
quantifies the information inherent in a non-detection of the stochastic background, and only depends on the search sensitivity $\sim P_{n1}, P_{n2}$. %
The relative information, Eq.~\eqref{eq:I_logOmegaref}, is then zero as expected for vanishingly small signals, as in this case, if the stochastic background is undetectable, then we have infinite fractional uncertainty on its size. %
This makes sense qualitatively, and though the Fisher formalism is only strictly applicable in the strong signal limit, it provides the Cramer-Rao lower bound on the variance of the estimator. %
The term $B(f)$, on the other hand, quantifies the information contributed by a measurable stochastic signal. {\color{black}The Fisher information $F^{\rm CC}_{\Omegaref}$ itself scales intuitively with the size of the stochastic background: given the negative sign of $B(f)$, as $\OmegaGW$ grows $F^{\rm CC}_{\Omegaref}$ \textit{decreases}. However, the Fisher information $F^{\rm CC}_{\log\Omegaref}$ can in principle increase or decrease, depending on which term dominates Eq.~\eqref{eq:I_logOmegaref}. That is, as $\OmegaGW$ is increased, absolute uncertainty on $\delta \Omegaref$ grows, but fractional uncertainty  $\delta \Omegaref/\Omegaref$ decreases as long as we are in the weak-signal approximation\footnote{\color{black}This may be seen qualitatively by observing that, in Eq.~\eqref{eq:stoch-total-info-absolute}, the $B(f)\Omega_f$ term can compete with the $A(f)$ term when $\Omega_f$ is of the order of the noise terms $P_{n, i}$.}.} %
Finally, information increases as observing time grows. %
At fixed analysis segment length $T_{\rm seg}$, the number of segments grows linearly with time and ${F}^{\rm CC}_{\Omegaref}\propto N_{\rm seg}$. This is expected as the variance of the measured stochastic field power scales as $1/T_{\rm obs}$~\cite{Romano:2016dpx}. %

\begin{figure}
    \centering
    \includegraphics[width=0.5\textwidth]{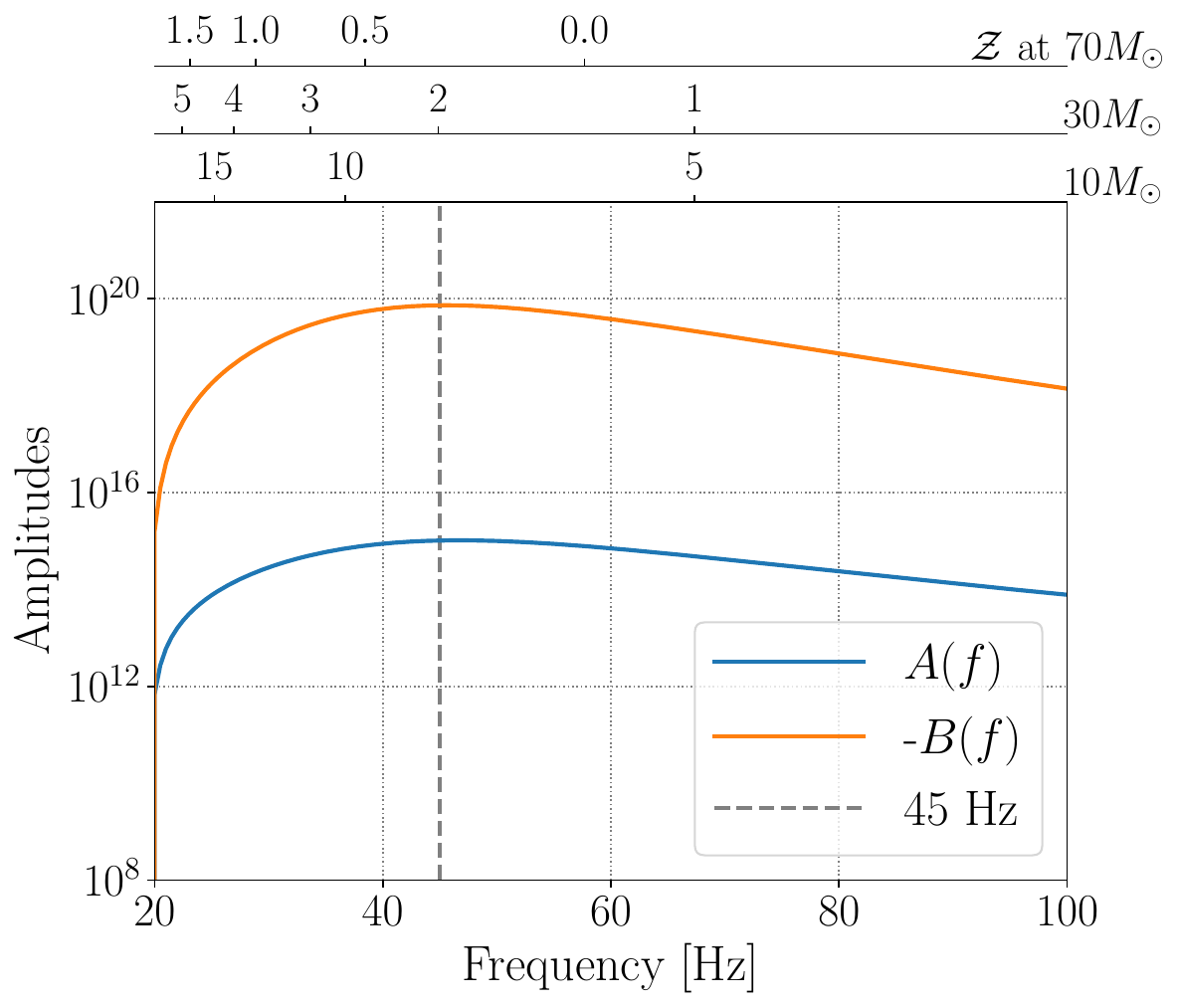}
    \caption{$A$ and $B$ spectra from Eq.~\eqref{eq:stoch-total-info-absolute} as a function of frequency. The top x-axes mark the redshift $\cal Z$ at which equal-component-mass binaries merge emitting at the frequency in the lower $x$-axis for different binary masses. The frequency corresponding to maximum stochastic sensitivity, $45$\,Hz, is marked with a grey dashed line.
    }
    \label{fig:A_B_spectra}
\end{figure}
{\color{black}Figure~\ref{fig:A_B_spectra} shows $A(f)$ and $-B(f)$, for the two-detector network of Advanced LIGO detectors, each with design sensitivity~\cite{design_sensitivity_curve} assuming 1 year of data. As seen in Eqs.~\eqref{eq:Af} and~\eqref{eq:Bf}, the functional form of these spectra depend on both the overlap reduction function and the individual detector sensitivities. % 
The 5 orders of magnitude between the two curves implies that for typical values of the background amplitude ($\Omega_f < 10^{-8}$) and at this sensitivity the $A(f)$ term dominates $F^{\rm CC}(\Omega_f)$, verifying the weak-signal approximation. %
In the case in question, the majority of the information is gathered at $\sim 45$\,Hz, where $A(f)$ is maximum for this specific configuration\footnote{\color{black} As discussed, the spectral shapes of $A(f)$ and $B(f)$ will necessarily vary for different detector pairs. 
In general, the larger the separation between detectors, the lower we expect these functions to peak in frequency, due to the form of the overlap reduction function~\cite{Romano:2016dpx}.}.} %
As sources at cosmological distances are redshifted, this "most informative" frequency can be translated into a "most informative" redshift, given source mass. %
At the top of Fig.~\ref{fig:A_B_spectra} we have added axes with the ``merger redshift'' $\cal Z$ 
at which equal-mass binaries of different masses merge, at the corresponding frequency on the $x$-axis. %
We select 3 component mass values for this example, $70\,M_\odot$, $30\,M_\odot$, $10\,M_\odot$, which we reprise in Sec.~\ref{sec:information} to calculate the Fisher information. %
Binaries with larger mass merge at lower frequencies, and conversely binaries with smaller mass contribute to lower frequencies only when highly redshifted. For example, a binary composed of two $70\,M_\odot$ black holes merges at around $\sim 60$Hz at $z=0$, and can contribute to lower frequencies in the band ($20< f < 60$Hz) when redshifted by $z<1.5$. On the other hand, a $10\,M_\odot$ equal-mass binary system can only contribute to a signal at frequencies $<60$Hz if redshifted by $z>6$. This implies that, in general, that $\Omega(f)$ at peak sensitivity $\sim 45$Hz is larger for higher-mass binary populations.

Finally, we calculate the Fisher matrix for the $2d$ extension to the cross-correlation search with parameters $R_0$ and $\zmax$, %
\begin{equation}\label{eq:I_R0_zmax}
    {\bm F}^{\rm CC}_{R_0, \zmax} \equiv   \begin{pmatrix}
        F^{\rm CC}_{R_0R_0} & F^{\rm CC}_{ R_0\zmax}\\
        F^{\rm CC}_{\zmax R_0} & F^{\rm CC}_{\zmax\zmax} 
    \end{pmatrix} \,.
\end{equation}
Each of the Fisher term here can be related to the Fisher matrix in $\Omegaref$ using the chain rule. The first diagonal term is
\begin{align}\label{eq:I_R0_CC}
    F^{\rm CC}_{R_0R_0} &= -\left\langle\frac{\partial^2\ln p}{\partial R_0^2}\right\rangle\nonumber= \qty(\frac{\Omegaref}{R_0})^2 \sum_f E^2(f) F^{\rm CC}_{\Omega_{f}}\nonumber\\
    &\equiv  R_0^{-2}F^{\rm CC}_{\log \Omegaref}\,,
\end{align}
using the fact that 
\begin{equation}
    \frac{\partial\Omega_f}{\partial R_0} \equiv \frac{\Omegaref}{R_0} E(f)\,,
\end{equation}
as per Eqs.~\eqref{eq:stoch-energy-density} and~\eqref{eq:Omega_ref}. %

The second diagonal term is similarly derived as
\begin{align}\label{eq:I_zmax_CC}
    F^{\rm CC}_{\zmax\zmax} &= -\left\langle\frac{\partial^2\ln p}{\partial \zmax^2}\right\rangle= \sum_f \qty(\Omega'_f)^2 F^{\rm CC}_{\Omega_{f}}\,,
\end{align}
where
\begin{equation}
\Omega'_{f} \equiv \frac{\partial\Omega_f}{\partial \zmax}\,,
\end{equation}
is the integrand of Eq.~\eqref{eq:stoch-energy-density} evaluated at $\zmax$.

Finally, the off-diagonal terms are
\begin{align}\label{eq:I_R0zmax_CC}
    F^{\rm CC}_{R_0\zmax}=F^{\rm CC}_{\zmax R_0} &= -\left\langle\frac{\partial^2\ln p}{\partial R_0\partial \zmax}\right\rangle\nonumber\\
    &= \qty(\frac{\Omegaref}{R_0}) \sum_f  E(f)\Omega'_f F^{\rm CC}_{\Omega_{f}}\,.
\end{align}
%

%------------------------------------------------------------------------------------
%
%
\subsection{Template-Based Search}

The template-based search likelihood,  Eq.~\eqref{eq:xi-likelihood-B}, depends on the redshift distribution parameters $R_0$ and $\zmax$ explicitly through the $\xi(R_0,\zmax)$ parameter, Eq.~\eqref{eq:xi-helper}, and implicitly through the Bayes factors $b_i(\zmax)$, Eq.~\eqref{eq:bi}, which depend on prior distributions on source parameters that in turn rely on $\zmax$. %
The likelihood is then re-written as
\begin{align}
    \label{eq:xi-likelihood-final}
    &p^{\rm TB}(\{d_i\}|R_0,\zmax) \nonumber \\
    &= \prod_i p(d_i | \mathcal{N}_i) \left[ \xi(R_0,\zmax) b_i (\zmax)  + (1-\xi(R_0,\zmax)) \right]\,,
\end{align}
and the Fisher matrix is
\begin{equation}\label{eq:I_matrix_TB}
    {\bm F}^{\rm TB}_{R_0,\zmax} \equiv   \begin{pmatrix}
        F^{\rm TB}_{\rm R_0R_0} & F^{\rm TB}_{\rm R_0\zmax}\\
        F^{\rm TB}_{\rm \zmax R_0} & F^{\rm TB}_{\rm \zmax\zmax} 
    \end{pmatrix} \,.
\end{equation}
Starting with the diagonal term in $R_0$ and writing $p^{\rm TB}\equiv p^{\rm TB}(\{d_i\}|R_0,\zmax)$, $\xi=\xi(R_0,\zmax)$, $b_i=b_i(\zmax)$ for conciseness, we find
\begin{align}
F^{\rm TB}_{R_0R_0}
    &= \left\langle - \frac{\partial^2}{\partial R_0^2} \ln p^{\rm TB}\right\rangle \nonumber\\
    &= -\left\langle \frac{\partial \xi}{\partial R_0}\frac{\partial}{\partial \xi} \left(
        \frac{\partial \xi}{\partial R_0}\frac{\partial}{\partial \xi} \ln p^{\rm TB}
        \right)\right\rangle \,.
\end{align}
The brackets signify ensemble averaging over data realizations, which we interpret in practice as utilizing all available data and evaluating the derivatives at the maximum of the likelihood. Since the first derivative vanishes at the maximum we get
\begin{equation}
    F^{\rm TB}_{R_0R_0}= - \qty(\frac{\partial \xi}{\partial R_0})^2 \left\langle \frac{\partial^2 \ln p^{\rm TB} }{\partial \xi^2} \right\rangle\,.
\end{equation}
Substituting Eq.~\eqref{eq:xi-likelihood-final} we obtain 
\begin{equation}\label{eq:I_R0_TB}
F^{\rm TB}_{R_0R_0}
    =  {\cal A}^2 \sum_i \frac{ (b_i - 1)^2}{(1+\xi(b_i - 1))^2} \,,
\end{equation} 
where 
\begin{equation}
    {\cal A} \equiv \frac{\partial \xi}{\partial R_0} = {\cal I}(\zmax) \tau e^{-R_0 {\cal I}(\zmax)\tau}\,= {\cal I}(\zmax) \tau (1-\xi)\,,
\end{equation}
where ${\cal I}(z_{\rm max})$ is defined as in Eq.~\eqref{eq:R0-helper}.
The Fisher matrix diagonal term in $\zmax$ is

\begin{equation}\label{eq:I_zmax_TB}
    F_{\zmax\zmax} =  \sum_i \frac{\qty(\xi'(b_i-1))^2 - 2 \xi' b_i' + \qty(\xi b_i')^2}{\qty(1+\xi(b_i - 1))^2} \,,
\end{equation}
where
\begin{align}
    b' &\equiv \frac{\partial b}{\partial \zmax}\,, \\ 
    \xi' &= \frac{\partial \xi}{\partial \zmax} = \tau R_0 {\cal I'}(\zmax) (1-\xi) \,.
\end{align}
Finally, the off-diagonal terms are 
\begin{align}
\label{eq:I_R0zmax_TB}
    F_{R_0 \zmax}&=F_{\zmax R_0} \nonumber\\
    &=  {\cal A} \sum_i\frac{\xi' (b_i-1)^2 - b_i'}{\qty(1+\xi(b_i - 1))^2}\,.
\end{align}
%

%%%%%%%%%%%%%%%%%%%%%%%%%%%%%%%%%%%%%%%%%%%%%%%%%%%%%
\section{Sensitivity reach}\label
{sec:I-study}
%%%%%%%%%%%%%%%%%%%%%%%%%%%%%%%%%%%%%%%%%%%%%%%

We quantify the dependence of the Fisher information on the merger rate redshift distribution with simulated populations. 
We adopt a constant normalised merger rate density $r(z)$ such that the merger rate $R(z)$ is uniform in comoving volume, a local merger rate of $R_0=30\, {\rm Gpc}^{-3}{\rm y}^{-1}$~\cite{PhysRevX.13.011048}, and vary $\zmax$ such that
\begin{equation}\label{eq:Rz_info_study}
    R(z) =  \left\{
\begin{array}{ll}
      \frac{R_0}{1+z} \frac{dV_c}{dz} & z\leq \zmax \\
      0, & z > \zmax \\
\end{array} 
\right. \,.
\end{equation}
We consider three distinct populations, made up of equal-mass, non-spinning black hole binaries with source-frame masses of $10 \,M_\odot$, $30 \,M_\odot$, and $70 \,M_\odot$, respectively. We calculate the Fisher information for each population at varying $\zmax$, over one year of data from a network of two LIGO detectors at design sensitivity~\cite{design_sensitivity_curve}.

%%%%%%%%%%%%%%%%%%%%%%%%%%%%%%%%%%%%%%%%%

%
\begin{figure*}[ht]
\centering
\includegraphics[width=\textwidth]{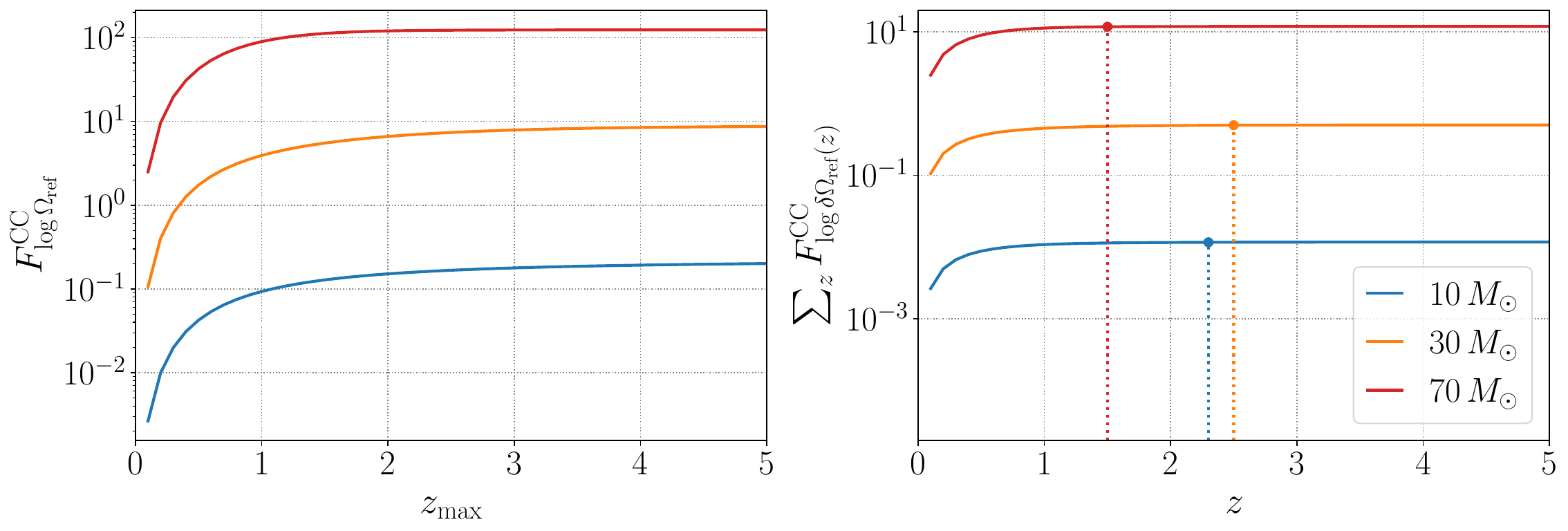}
\caption{Left: Single-parameter Fisher matrix $F^{\rm CC}_{\log\Omegaref}$, Eq.~\eqref{eq:I_logOmegaref}, calculated for varying $\zmax$, for each equal-mass binary population considered. %
Right: Cumulative sum of the information $F^{\rm CC}_{\log\delta\Omegaref(z)}$ contributed from binaries in a redshift shell $[z,z+\delta z]$ as a function of $z$ for a Universe with binaries up to $\zmax=5$ and for each mass. %
The dots indicate the redshift at which 99\% of information is accumulated. 
Both plots indicate that the information saturates at a mass-dependent redshift $\sim$2.}
\label{fig:I_Omega_CC}
\end{figure*}
\begin{figure*}[t]
\centering
\includegraphics[width=\textwidth]{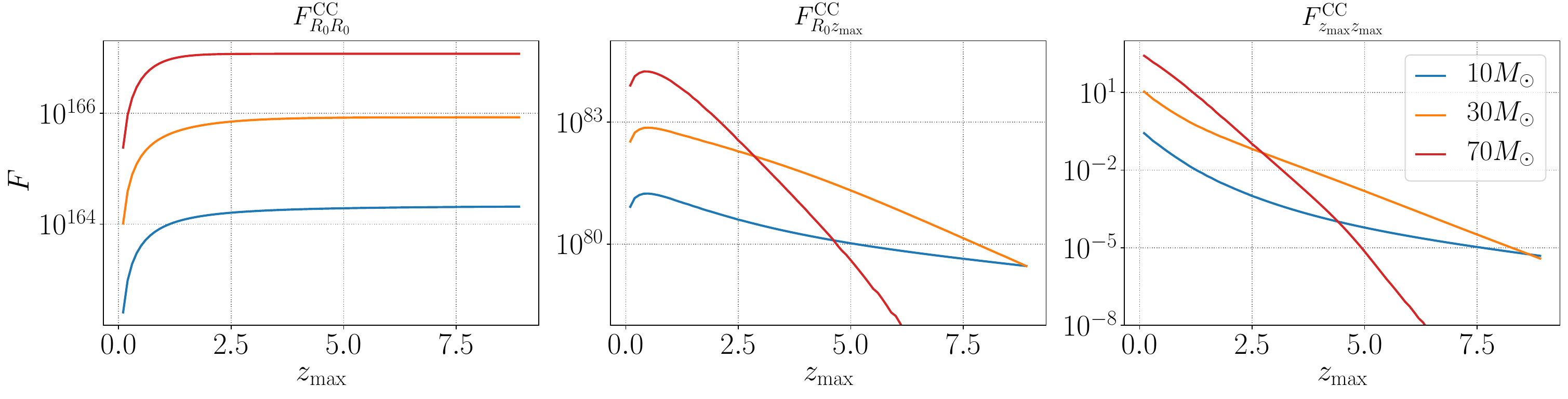}
\caption{Fisher matrix terms for the two-parameter, $2d$ version of the cross-correlation search for $R_0, \zmax$ for different equal-mass populations. We consider Universes with varying $\zmax$ and plot the Fisher terms as a function of $\zmax$. }
\label{fig:I_matrix_CC}
\end{figure*}
%

%%%%%%%%%%%%%%%%%%%%%%%%%%%%%%%%%%%%%%%
\subsection{Cross-Correlation Search}

For the cross-correlation search, both the $1d$ and the $2d$ Fisher matrices in Eqs.~\eqref{eq:ICC_logOm} and~\eqref{eq:I_R0_zmax} can be calculated analytically. %
The spectrum $\OmegaGW(f)$ is calculated for each population and at each $\zmax$ using Eq.~\eqref{eq:stoch-energy-density}. % 
The emitted energy spectrum $dE_s/df_s$ is a function of source-frame chirp mass ${\cal M}_c$~\cite{Zhu_2011}, %  
\begin{equation}\label{eq:dE_dfs}
    \frac{dE_s}{df_s} = \frac{(G\pi)^{2/3} {\cal M}_c^{5/3}}{3} {\cal E}(f_s)\,,
\end{equation}
where the function ${\cal E}(f_s)$ may be found, e.g., in~\cite{Zhu_2011}. %
During the inspiral phase, i.e., for frequencies lower than the merger frequency, ${\cal E}(f_s) \approx f_s^{-1/3}$, implying $\OmegaGW(f)\propto f_s^{2/3}$, which is generally a good approximation to the low-frequency background spectral shape. %
To model the background spectrum at higher frequencies, we adopt an analytical inspiral-merger-ringdown approximation for ${\cal E}(f_s)$~\cite{Zhu_2011,Ajith:2007kx}. %

We start by considering the single-parameter, $1d$ case. We compute the Fisher matrix in $\log\Omegaref$ of Eq.~\eqref{eq:I_logOmegaref} varying the cutoff $\zmax$ in the computation of $\Omegaref$ and plot $F^{\rm CC}_{\log \Omegaref}$ as a function of $\zmax$ in the left panel of Fig.~\ref{fig:I_Omega_CC}. %
This plot illustrates how the Fisher matrix (equivalently, the Fisher information as we are in $1d$) increases with $\zmax$ as further distant binaries contribute to the total $\Omegaref$.
The information plateaus at varying $\zmax$, depending on the mass, reaching higher values for higher masses as expected. % 
The $\zmax$ at which $F^{\rm CC}_{\log \Omegaref}$ plateaus is a combination of the redshift at which the binaries emitting at the most sensitive frequency merge ($\cal Z$ in Fig.~\ref{fig:A_B_spectra}), and the redshift at which binaries no longer appreciably contribute to $\Omegaref$. 
For example, for $z_{\rm max}=5$, 99\% of the background amplitude is accumulated from binaries within $z<2.9$, independently of mass.%

To understand how binaries in each redshift bin contribute to the total information, we consider the case of a Universe with binaries up to $\zmax=5$.
We calculate the contribution of a redshift shell $[z,z+\delta z]$ to the background  $\delta\OmegaGW (z) = \OmegaGW(z+\delta z) - \OmegaGW(z)$ and plug this into the information Eq.~\eqref{eq:I_logOmegaref} to obtain the information contribution of that shell $F^{\rm CC}_{\log\delta\Omegaref(z)}$. 
The right panel of Fig.~\ref{fig:I_Omega_CC} shows the cumulative sum of $F^{\rm CC}_{\log\delta\Omegaref(z)}$, i.e., the cumulative information in $\log\Omegaref$, for each mass. 
We pinpoint on the plot the redshift at which 99\% the information has been accumulated: $z^{99\%}_{10M_\odot} = 2.2$, $z^{99\%}_{30M_\odot} = 2.5$, $z^{99\%}_{70M_\odot} = 1.5$. This is evidently highly sensitive on the mass.

We now turn to the two-parameter, $2d$ information matrix shown in Eq.~\eqref{eq:I_R0_zmax} for the $R_0$ and $\zmax$ parameters. Similarly to the single-parameter case above, we compute the individual matrix components for varying $\zmax$ and plot them in Fig.~\ref{fig:I_matrix_CC} as a function of $\zmax$. %
The diagonal $F^{\rm CC}_{R_0 R_0}$ term has the same form as $F^{\rm CC}_{\log\Omegaref}$, up to an $R_0^2$ rescaling, as seen in Eq.~\eqref{eq:I_R0_CC}, and may be similarly interpreted. %
The diagonal $F^{\rm CC}_{\zmax \zmax}$ term, Eq.~\eqref{eq:I_zmax_CC}, monotonically decreases as it is dominated by the $\Omega'(f)$ term summed over frequencies, which decreases as less and less $\Omega(f)$ is accumulated at higher redshifts. % 
Stated differently, information about $\zmax$ is provided by binaries close to $\zmax$, which have lower signal-to-noise ratio (SNR) as $\zmax$ increases.
The cross term $F^{\rm CC}_{R_0\zmax}$, Eq.~\eqref{eq:I_R0zmax_CC}, encodes correlations between $R_0$ and $\zmax$ which decrease as $\zmax$ increases. %

%%%%%%%%%%%%%%%%%%%%%%%%%%%%%%%%%%%%%%%%%
\subsection{Template-Based Search}\label{sec:TB_results}

\begin{figure*}[t]
\centering
\includegraphics[width=\textwidth]{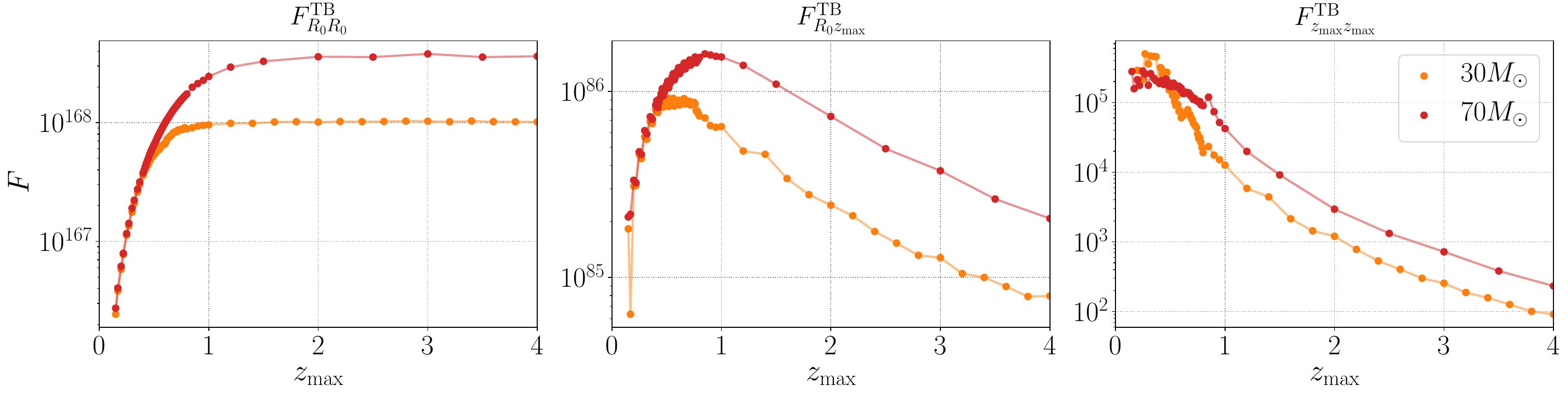}
\caption{Similar to Fig.~\ref{fig:I_matrix_CC} but for the template-based search. 
Dots indicate the $\zmax$ values used for the numerical realizations and lines correspond to interpolation. 
%Curves have further been smoothed with a Gaussian filter to reduce numerical noise at $\zmax<1.5$. 
Ringing in the curves at $\zmax<1$ is due to the low number of events at these redshifts. %
The Fisher terms display qualitatively similar trends as the equivalent results for the cross-correlation search in Fig.~\ref{fig:I_matrix_CC}.}
\label{fig:I_matrix_TB}
\end{figure*}
\begin{figure*}[t]
\centering
\includegraphics[width=\textwidth]{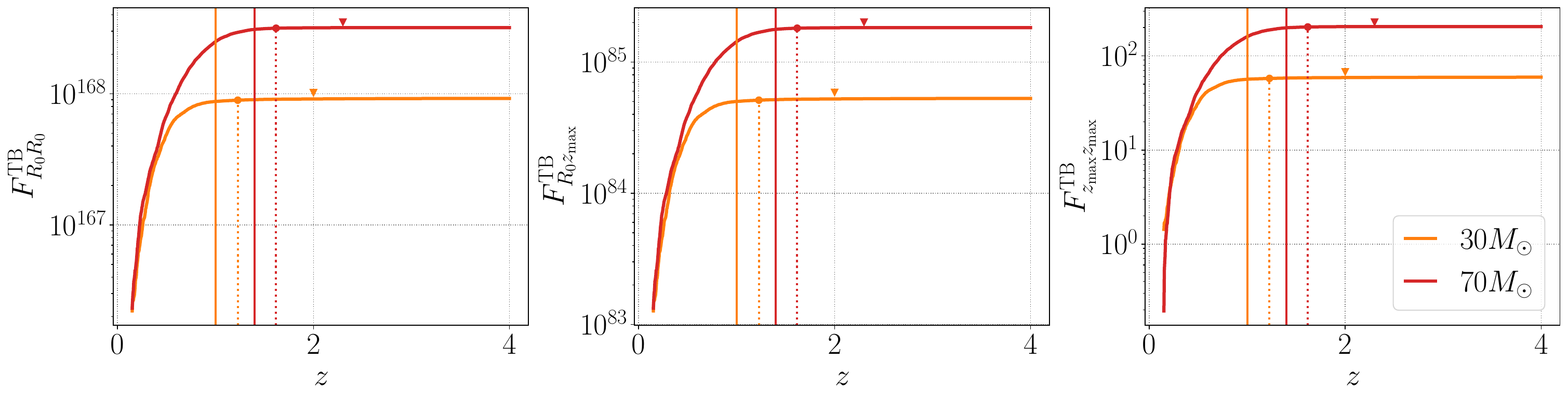}
\caption{Similar to the right panel of Fig.~\ref{fig:I_Omega_CC} but for the template-based search and all terms of the $2d$ Fisher. Here we consider the $\zmax=4$ case. 
Dots again indicate the redshift up to which 99\% of each term has been accumulated.
Vertical solid lines indicate the redshift $z^{\rm res}$ at which events have on average SNR $=8$, for each mass. The plateau in each Fisher term occurs slightly beyond this SNR threshold. 
Arrow markers indicate the redshift of the maximum SNR horizon cutoff for each mass considered.
}

\label{fig:F_cumu}
\end{figure*}

The information content of the template-based search as outlined in Eqs.~\eqref{eq:I_matrix_TB},~\eqref{eq:I_R0_TB},~\eqref{eq:I_zmax_TB},~\eqref{eq:I_R0zmax_TB} needs to be computed numerically. %
In a real search, parameter estimation needs to be performed on all time segments within a given dataset in order to marginalize over the $15$ parameters of quasicircular binaries and calculate the Bayes factor $b_i$ for each segment $i$. %
To avoid the prohibitive computational cost of such an analysis, here we consider a simplified toy model where the only inferred parameter is the distance of the binary, while all other parameters are known. %
This simplification will artificially boost the sensitivity of the search, resulting in unrealistically high estimates for the total Fisher information. However, we adopt this approximation as we are more interested in the Fisher dependence with redshift than its absolute value. %

We proceed to compute the information matrix numerically for different $\zmax$ values up to $\zmax = 4$ and for equal-mass populations of binaries with component masses 30$M_\odot$ and $70\,M_\odot$.\footnote{We omit the lowest mass, $10\,M_\odot$, as preliminary calculations were dominated by numerical noise.} %
For each $\zmax$ we draw binaries up to that redshift and simulate one year of data, which we divide into 
 $N_{\rm seg}=7,889,400$ segments of length $\tau=4$s. %
In practice, given $R_0$ and $r(z)$, we calculate the number of expected events during a year of observation, $N^{\rm max}_{\rm ev}=16,470$, for the largest $\zmax$ simulation. %
Assuming that there is at most one event per data segment, $N^{\rm max}_{\rm ev}$ is then the number of segments that contain a signal for that realization. 
For each mass population and $\zmax$, we generate data by considering the subset of events with $z<\zmax$, resulting in $N_{\rm ev}$ segments with events. %
We then calculate the Bayes factor $b_i$ of the search for each segment assuming a redshift prior of $r(z)$ with the appropriate $\zmax$ as detailed in App.~\ref{app:bfs}. 

This procedure leaves the vast majority of segments without signals which still need to be summed over, e.g., in Eq.~\eqref{eq:I_R0_TB}.
To estimate the noise segment contribution, we employ the {\tt Bilby} library~\cite{Ashton:2018jfp} and simulate $N_{\rm noi}$ realisations of Gaussian noise. %
For each segment $i$, we numerically compute the Bayes factor, $b_i$, and its derivative with respect to $\zmax$, $b'_i$, with the same prior as the signal segments, as described in App.~\ref{app:bfs}. %
%In practice, to gain a handle on the population variance for the signal realisations, we average the signal contribution to the information over 10 independent realisations of signal and noise, for each mass. %
To reduce the computational cost, we restrict $N_{\rm noi}=78,800$ which are then re-used to obtain the desired number of noise segments $N_{\rm seg}-N_{\rm ev}$. %
This procedure could potentially cause artifacts in the noise contribution to the information, but we assume these will not dominate the total information, as the information contributed by the noise segments will be much smaller than that contributed by segments containing events, and we expect $N_{\rm noi}$ to be sufficiently large to capture the variability in noise-segment Bayes' factors. 

The information matrix terms for each realization are shown in Fig.~\ref{fig:I_matrix_TB} as a function of $\zmax$.
These terms exhibit qualitatively similar trends as Fig.~\ref{fig:I_matrix_CC} for the cross-correlation search, albeit with some numerical noise. %
The $F^{\rm TB}_{R_0 R_0}$ term increases monotonically to a mass-dependent plateau, as the information on the local merger rate is concentrated at lower redshifts. 
The plateau value varies with mass as higher mass binaries may be detected further away.
The $F^{\rm TB}_{R_0 \zmax}$ and $F^{\rm TB}_{\zmax \zmax}$  terms peak at different $0<\zmax<1.5$ redshifts, depending on the interplay between the Bayes factor (drastically higher at lower redshifts) and the number of events per redshift shell. %

We then restrict to the $\zmax=4$ realization and calculate the cumulative contribution of events to the Fisher terms as a function of redshift. 
This is similar to the right panel of Fig.~\ref{fig:I_Omega_CC} for the cross-correlation search, only here we consider all Fisher terms.
Moreover, since the template-based Fisher terms are already expressed as a sum over events, we simply restrict the sums to segments with events in that redshift bin. %
Results are shown in Fig.~\ref{fig:F_cumu}. %
The three $F^{\rm TB}$ terms plateau in redshift according to the mass. We consider the redshift value up to which 99\% of each term has been accumulated (indicated with a dotted line in Fig.~\ref{fig:F_cumu}):  $z^{99\%}_{30M_\odot}=1.2$, $z^{99\%}_{70M_\odot}=1.6$. %
These are the same for the three terms. %
Vertical solid lines denote the ``resolvability" cutoff, i.e., the redshift at which the average two-detector SNR is 8: $z^{\rm res}_{30M_{\odot}}=1.0$ and $z^{\rm res}_{70M_{\odot}}=1.4$. %
The resolvability cutoffs are close to the redshifts at which the terms plateau, with $<5\%$ of the contribution coming between $z^{\rm res}$ and $z^{99\%}$. %
For reference, the maximum redshift at which binaries with these masses can be observed at the chosen sensitivity are $z^{\rm max}_{30M_{\odot}}=2.0$ and $z^{\rm max}_{70M_{\odot}}=2.4$. % 
%Furthermore, approximately 0.3\% of resolved events lie beyond $z>z^{\rm res}$ \kc{we could show this by plotting the horizon as well per the comment above}. %

%%%%%%%%%%%%%%%%%%%%%%%%%%%%%%%%%%%%%%
\subsection{Comparison and Discussion}

\begin{figure*}[t]
\centering
\includegraphics[width=\textwidth]{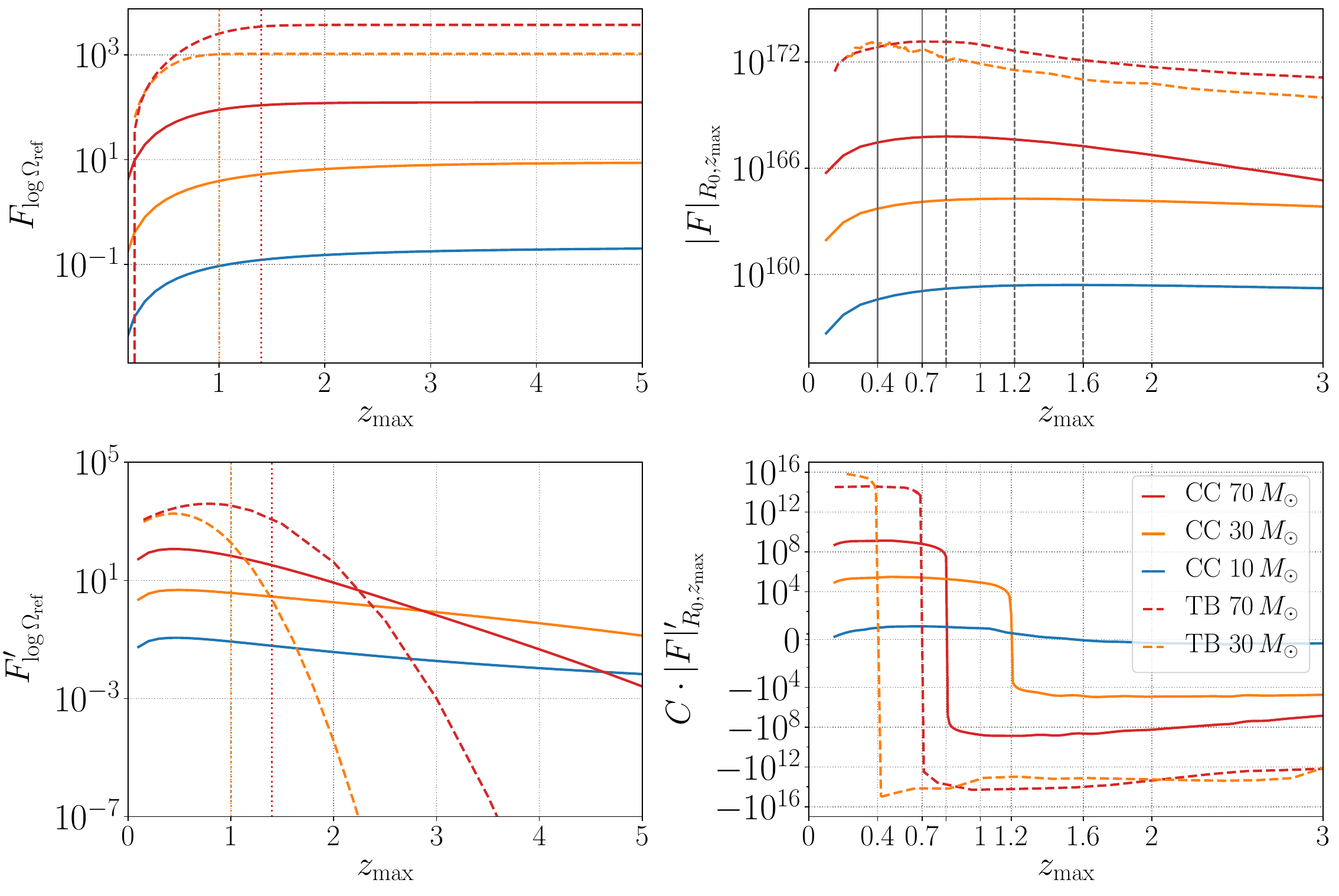}
\caption{
Top left: $F^{\rm CC}_{\log \Omegaref}$ and $F^{\rm TB}_{\log \Omegaref}$ for the cross-correlation and template-based searches respectively as a function of the maximum redshift the population extends to, $\zmax$, and different masses. Though the absolute value of the information is not comparable between searches, they both show a similar trend with $\zmax$, plateauing at some mass-dependen value for $\zmax\sim 1-2$.
Bottom left: derivatives of $F^{\rm CC}_{\log \Omegaref}$ and $F^{\rm TB}_{\log \Omegaref}$ with respect to $\zmax$, $F'_{\log \Omegaref}$. %
Dotted lines in the left column plots report the SNR $=8$ resolvability cutoffs discussed in the text. %
Right column: similar to the left panels but for the determinant of the $2d$ Fisher matrix in the case of the two-parameter search for $R_0$ and $\zmax$. %
In the latter plot, all curves have been multiplied by a constant $C=10^{-159}$ for visualization purposes. %
The values of $\zmax$ at which derivatives change sign for the Fisher matrix determinants of the two searches are marked on the x-axis, and are shown in grey solid  (dashed) lines for the template-based (cross-corelation) search results in the top panel.
}
\label{fig:I_det_comp}
\end{figure*}

We now compare the information content and sensitivity reach of the template-based and cross-correlation searches. Given the simplifications made to the calculation of the template-based Bayes' factors, a direct comparison of the total information accumulated by the two searches is not meaningful; in reality, the template-based search as formulated in~\cite{Smith:2017vfk, Smith:2020lkj} would require sampling and marginalizing over many more waveform and population parameters, which would impact the recovered information. %
Instead, we focus on comparisons of the redshift dependence of the information and specifically the relative contribution of different redshifts 
in the cases of single-parameter and two-parameter searches. %

Starting with the single-parameter case, we consider the current formulation of the template-based search that assumes a known rate distribution~\cite{Smith:2017vfk, Smith:2020lkj}.
We select $\log \Omegaref$ as the common parameter to compare searches with.
For the cross-correlation search, the relevant information is given in Eq.~\eqref{eq:I_logOmegaref} (expressed as the Fisher term, which is identical to the information in $1d$).
For the template-based search, we need to convert Eq.~\eqref{eq:I_R0_TB} from $R_0$ to $\log \Omegaref$.
Equations~\eqref{eq:stoch-energy-density} and~\eqref{eq:Omega_ref} imply that $R_0$ and $\Omegaref$ differ solely by a constant,
\begin{equation}
    \frac{\partial R_0}{\partial \Omegaref} \equiv \frac{R_0}{\Omegaref}\,.
\end{equation}
It is then straightforward to calculate the information in $\log \Omegaref$ from the information in $R_0$,
\begin{equation}
    F^{\rm TB}_{\log \Omegaref}=F^{\rm TB}_{\log R_0} =R_0^2 F^{\rm TB}_{R_0R_0}\,,
\end{equation}
where $F^{\rm TB}_{R_0R_0}$ is given in Eq.~\eqref{eq:I_R0_TB}.
The top panels of Fig.~\ref{fig:I_det_comp} show $F^{\rm TB}_{\log \Omegaref}$ and $F^{\rm CC}_{\log \Omegaref}$ as a function of $\zmax$ (top) and their derivatives with respect to $\zmax$ (bottom).\footnote{To avoid taking the numerical derivative of the noisy  $F^{\rm TB}_{R_0R_0}$ term shown in Fig.~\ref{fig:I_matrix_TB}, we fit it with an error function and take the derivative of the latter.}.

Starting with the top left panel, the template-based search information in $\log\Omegaref$ reaches ${\cal{O}}(10^2)$ larger values than the cross-correlation one at all redshifts. %
This result is qualitatively expected as a phase-coherent search is more sensitive than a power-based one especially when loud, resolved signals are concerned. 
However, the quantitative result is subject to the caveat about the simplicity of our template-based search implementation. 
More interesting, on the other hand, is the dependence with $\zmax$, as information in both searches saturates at $\zmax \sim 1-2$.
Increasing $\zmax$ means adding binaries at high redshift. 
Once the information saturates, adding more binaries at higher redshifts does not contribute information about the total energy density.
In other words, the gravitational-wave energy density is primarily informed from binaries at redshift $\lesssim1-2$. 

The bottom left panel of Fig.~\ref{fig:I_det_comp} further demonstrates this point through the derivative of the information with respect to $\zmax$.
In the template-based case, the derivative is larger at low redshifts $\lesssim 1$ before sharply turning over.
In the cross-correlation case, the derivative exhibits a more slight decline with $\zmax$.
This suggests that in the template-based search, local binaries contribute relatively more information than distant ones.
For reference, vertical dotted lines again show the resolvability cutoff, where the average signal SNR is 8. %
In both searches, the information trend is mass-dependent. As $\zmax$ increases, sources are redshifted to higher masses and lower frequencies. More massive sources are potentially redshifted out of the LIGO frequency band, while light sources are redshifted into the most sensitive band region. %
The stark difference in the mass-dependent trends between the searches is due to their different frequency sensitivity: the cross-correlation search includes an overlap reduction function that pushes the sensitivity to lower frequencies, while the template-based search is most sensitive to the central part of the LIGO sensitivity band $\sim 10^2$\,Hz~\cite{sensitivity_curves}.

Finally, we examine the two-parameter searches in the right panel of Fig.~\ref{fig:I_det_comp} by plotting the determinant of the $2d$ Fisher matrix from Eqs.~\eqref{eq:I_R0_zmax} and~\eqref{eq:I_matrix_TB} (top) and its derivative (bottom). %
Adding $\zmax$ as a parameter allows us to probe directly the ability of each search to infer the high-redshift population properties. %
As in the $1d$ case, the amplitude of the information is much higher for the template-based search across all $\zmax$ values; this conclusion is subject to the usual caveats about our simplified analysis.
More importantly, the derivatives (bottom right panel) in all cases peak at varying redshifts before sharply declining and becoming negative. For the cross-correlation search the derivatives start declining at $\hat{z}^{\rm CC}_{10M_\odot}=1.6$,  $\hat{z}^{\rm CC}_{30M_\odot}=1.2$, $\hat{z}^{\rm CC}_{70M_\odot}=0.8$, while the same numbers for the template-based search are $\hat{z}^{\rm TB}_{30M_\odot}=0.4$,  $\hat{z}^{\rm TB}_{70M_\odot}=0.7$. %
The fact that the derivatives decline at a higher redshift, again suggests that higher-redshift events have a larger impact relative to low-redshift events for the cross-correlation than the template-based search. %

Returning to the fact that the derivatives become negative after their initial peak, the template-based search results in a larger decline of the total information as $\zmax$ increases, i.e., more negative derivatives.
Assuming that information about $R_0$ is almost entirely coming from redshift $\lesssim 1-2$ sources, e.g., top panel of Fig.~\ref{fig:I_det_comp}, the decline in information comes from the fact that as $\zmax$ increases, it becomes less well measured.

%%%%%%%%%%%%%%%%%%%%%%%%%%%%%%%%%%%%%%%%%%
\section{Conclusions}\label{sec:conclusions}

In this paper, we compared the information content of two searches for the gravitational-wave background from the binary black holes: the cross-correlation and template-based searches. 
The two searches employ different methodologies as well as astrophysical assumptions about the black hole population.
As a consequence, they result in considerably different {\color{black}and thus complementary} sensitivities to the binary black hole population, in terms of both mass and redshift distribution. 
Specifically, the template-based search collects most of its information from binaries at lower redshift than the cross-correlation one.
In other words, detection of the stochastic background with the former probes binaries at lower redshifts than detection with the latter. %
{\color{black}Though the exact numerical results presented in Sec.~\ref{sec:I-study} depend on our specific simulated 
population, we expect this qualitative trend to be robust for most black hole masses measured to date. % 
However, for larger masses, where the signal is shorter in band and peaks at lower frequencies, this may no longer hold. %
Ideally, a detection with both searches would allow us to complement and expand our knowledge of the binary population as a function of both redshift and mass.
} %

We further clarified the astrophysical assumptions of each search and how these affect their sensitivity. 
At their simplest form, the template-based search targets the event rate while the cross-correlation search is formulated directly through the gravitational-wave energy density. 
While the latter can be directly measured, the former requires assumptions about the astrophysical distribution of black holes in the form of the priors that enter its Bayes factor calculation.
Extensions of the template-based search to simultaneously infer the redshift distribution with the event rate (as was done with the mass and spin distribution~\cite{Smith:2020lkj}) would likely reduce its sensitivity, but we argue that they are essential.
Assuming a known redshift distribution when detecting the stochastic background mixes information between low-redshift, resolvable events and high-redshift, unresolvable ones. %
{\color{black} To quantify this effect, we considered the case of fixed equal-mass binaries and compared the redshift cutoff at which the binaries have ${\rm SNR}=8$ on average with the redshift where the Fisher information saturates (see Fig.~\ref{fig:F_cumu} and related discussion). We found the two to be very close, and significantly smaller than the maximum redshift at which the same binaries may be observed; less than 6\% of the information is accumulated beyond the average resolvability cutoff.} %

{\color{black}In addition to the two search methods discussed in this paper, a hybrid search method~\cite{Lawrence:2023buo} targets the excess cross-correlated power assuming the intermittent Gaussian mixture-model of Eq.~\eqref{eq:xi-likelihood-A}. %
A similar information analysis dedicated to this novel search will be necessary to assess a detection. %
As this intermittent cross-correlated search relies on the combination of cross-correlated data in a mixture-model likelihood, we expect its capabilities and sensitivity to lie somewhere in between the cross-correlation and template-based searches described here.
}

O4 and future O5 data bring us closer to the detection of an astrophysical stochastic background.
At the same time,  searches for this background employ vastly different methodologies and assumptions. 
As these searches adapt their formulation, it remains essential to study the sensitivity reach of whichever search claims detection. %
%For example, the test presented in~\cite{Smith:2020lkj} to assess the ability to infer the cutoff parameter in the redshift distribution used (equivalent to $\zmax$ used in this paper) is a valuable starting point. % 
%Efforts such as ours and that of~\cite{Smith:2020lkj} will be necessary to validate and interpret a detection.
%KC: I moved this to the introduction, I'd rather we do not finish the paper on the note that someone else did something similar but we bring it up here for the first time. 

%%%%%%%%%%%%%%%%%%%%%%%%%%%%%%%%%%%%%%%%%%%%%%%%%%

%%%%%%%%%%%%%%%%%%%%%%%%%%%%%%%%%%%%%%%%%%%%%%%%%%
\acknowledgements

We thank Michele Vallisneri for useful discussions. 
AIR is supported by the European Union's Horizon 2020 research and innovation programme under the Marie Skłodowska-Curie grant agreement No 101064542, and acknowledges support from the NSF award PHY-1912594. %
TC is supported by the Eric and Wendy Schmidt AI in Science Postdoctoral Fellowship, a Schmidt Futures program.
KC acknowledges support from NSF Grant PHY-2110111 and the Sloan Foundation. %
WF is supported in part by the Simons Foundation.
The authors are grateful for computational resources provided by the LIGO Laboratory and supported by National Science Foundation Grants PHY-0757058 and PHY-0823459. 
Software packages employed: {\tt numpy}~\cite{harris2020array}, {\tt scipy}~\cite{2020SciPy-NMeth}, {\tt gwpy}~\cite{gwpy}, {\tt Bilby}~\cite{Ashton:2018jfp}, {\tt matplotlib}~\cite{Hunter:2007}, {\tt seaborn}~\cite{Waskom2021}.

\appendix

%%%%%%%%%%%%%%%%%%%%%%%%%%%%%%%%%
\section{Bayes factor calculation for the template-based search}\label{app:bfs}

The template-based search Fisher calculation requires Bayes factors $b_i$ and their derivatives $b'_i$ with respect to $\zmax$ for each segment $i$. %
The data are given by the sum of the signal strain $h_i$ plus a Gaussian noise realisation $n_i$,
\begin{equation}
    d_i(f) = h_i(f) + n_i(f)\,.
\end{equation}
In the case of noise-only segments, where no signal is injected, $h_i\equiv 0$. %
For segments with a signal, the strains $h_i$ are obtained by simulating \textsc{IMRPhenomD}~\cite{Khan:2015jqa} waveforms in a LIGO interferometer using the {\tt Bilby} package~\cite{Ashton:2018jfp} injection routine, assuming Gaussian noise with design sensitivity~\cite{sensitivity_curves}. %
The distances used for the injections are drawn from the normalized distance merger rate distribution $p(z)$ shown in Eqs.~\eqref{eq:p_of_z} and~\eqref{eq:Rz_info_study} with a $\zmax=4$. %
All other binary parameters remain fixed: we assume zero-spin, face-on binaries at $0^\circ$ right ascension, $0^\circ$ declination. %
These correspond to the best configuration which maximizes the binary SNR;
to account for the average binary inclination and sky position, we rescale the strain by an angle-average factor of $2.2648$~\cite{
Finn:1992xs, OShaughnessy:2009szr}. %

To calculate $b_i$, we take the same (uniform-in-comoving-volume) distance distribution $p(z)$ used to simulate the BBH population as our distance prior. %
We construct a \textit{single-parameter model} in redshift $z$ for the signal, which we will marginalize over, and generate a set of frequency-domain templates $h(f, z)$ for varying $z$. %
Planck15 cosmology~\cite{Planck:2015fie} is assumed to relate redshift and luminosity distance, where necessary. %
All other parameters in the templates are fixed to match the injections detailed above. %
We generate 3 template sets, one for each fixed source-frame binary mass: 10, 30, and 70\,$M_{\odot}$. %

The Bayes factor for data segment $i$~\cite{Thrane_Talbot_2019}
\begin{equation}\label{eq:bi}
    b_i(\zmax) = \frac{\int_0^{\zmax}dz\, p(z) \, p (d_i | z, S)}{\int_0^{\zmax}dz\, p(z)\, p (d_i | N)}\,;
\end{equation}
here $p(d_i|z, S)$ and $p(d_i|N)$ are the signal and noise likelihood, respectively, which are defined as
\begin{equation}
    \begin{aligned}
%        p(d_i|z, S) &= \frac{1}{2\pi\sigma^2} e^{-\frac{1}{2}\qty(d_i-h(z)|d_i-h(z))}\,, \\
%        p(d_i|N) &= \frac{1}{2\pi\sigma^2} e^{-\frac{1}{2}\qty(d_i|d_i)}\,, 
        p(d_i|z, S) &\sim e^{-\frac{1}{2}\qty(d_i-h(z)|d_i-h(z))}\,, \\
        p(d_i|N) &\sim e^{-\frac{1}{2}\qty(d_i|d_i)}\,, 
    \end{aligned}
\end{equation}
where the brackets $(\,|\,)$ denote the sensitivity-weighted inner product
\begin{equation}
    (a|b) = 4 {\rm Re}\qty[ \int_0^\infty df \frac{a(f) \cdot b^\star(f)}{P(f)} ]\,,
\end{equation}
where $P(f)$ is the one-sided detector power spectral density. %
Expanding out the inner products, simplifying terms and setting the normalization of $p(z)$ to 1, this reduces to
\begin{equation}\label{eq:bi2}
    b_i(\zmax) = \int_0^{\zmax}dz\, p(z) e^{-\frac{1}{2} \qty(h(f, z)|h(f, z)) + \qty(d_i(f)|h(f, z))}\,.
\end{equation}
%The first term in the exponential of Eq.~\eqref{eq:bi} gives rise to a constant for all segments.
The derivative of the Bayes factor with respect to $\zmax$ is the integrand of Eq.~\eqref{eq:bi2} evaluated at $\zmax$,
\begin{align}\label{eq:biprime}
    &b'_i(\zmax) = p(\zmax) \times \nonumber\\
    &\Bigl( e^{-\frac{1}{2} (h(f, z)|h(f, z))|_{\zmax} + (h(f, z)|d_i(f))|_{\zmax}}- b_i(\zmax)\Bigr)\,.
\end{align}

To account for the use of 2 detectors, we substitute $d_i$ above with a 2$d$ data vector with identical entries, $\bm{d}_i=(d_i, \,d_i)$, assuming we have two identical co-aligned and co-located detectors. %
To further simplify our calculation, we use two limits to calculate Bayes factors for signal and noise segments, respectively. %
In the presence of a signal, we assume this dominates the likelihood calculation such that the 2-detector Bayes factor reduces to
\begin{equation}
    \begin{aligned}
    b^{\text{2-det}}_i(\zmax) &\approx \int_0^{\zmax}dz\, p(z) e^{-\qty(h(f, z)|h(f, z)) +2 \qty(h_i(f)|h(f, z))}\,\\
    &= e^2\, b_i(\zmax)\,.
    \end{aligned}
\end{equation}
This approximation leads to optimistic estimates of signal vs. noise Bayes factors. %
Conversely, in the presence of noise the 2-detector Bayes factor is
\begin{equation}
\begin{aligned}
    b^{\text{2-det}}_i&(\zmax) = \int_0^{\zmax}dz\, p(z) e^{- \qty(h(f, z)|h(f, z))} \\
    &\qquad\qquad\qquad\times  e^{\qty(n_{1, i}(f)+n_{2, i}(f)|h(f, z) )}\,\\
    &=\int_0^{\zmax}dz\, p(z) e^{- \qty(h(f, z)|h(f, z)) + \sqrt{2}\qty(n_{i}(f)|h(f, z) ))}\,.
\end{aligned}
\end{equation}
where in the final equality we assume the two noise realisations are drawn from the same distribution. In practice, we draw $n_1$ and $n_2$ individually as this is more convenient in our code implementation.

\bibliography{bib}

%apsrev4-2.bst 2018-12-27 (MD) hand-edited version of apsrev4-1.bst
%Control: key (0)
%Control: author (72) initials jnrlst
%Control: editor formatted (1) identically to author
%Control: production of article title (-1) disabled
%Control: page (0) single
%Control: year (1) truncated
%Control: production of eprint (0) enabled
\begin{thebibliography}{53}%
\makeatletter
\providecommand \@ifxundefined [1]{%
 \@ifx{#1\undefined}
}%
\providecommand \@ifnum [1]{%
 \ifnum #1\expandafter \@firstoftwo
 \else \expandafter \@secondoftwo
 \fi
}%
\providecommand \@ifx [1]{%
 \ifx #1\expandafter \@firstoftwo
 \else \expandafter \@secondoftwo
 \fi
}%
\providecommand \natexlab [1]{#1}%
\providecommand \enquote  [1]{``#1''}%
\providecommand \bibnamefont  [1]{#1}%
\providecommand \bibfnamefont [1]{#1}%
\providecommand \citenamefont [1]{#1}%
\providecommand \href@noop [0]{\@secondoftwo}%
\providecommand \href [0]{\begingroup \@sanitize@url \@href}%
\providecommand \@href[1]{\@@startlink{#1}\@@href}%
\providecommand \@@href[1]{\endgroup#1\@@endlink}%
\providecommand \@sanitize@url [0]{\catcode `\\12\catcode `\$12\catcode
  `\&12\catcode `\#12\catcode `\^12\catcode `\_12\catcode `\%12\relax}%
\providecommand \@@startlink[1]{}%
\providecommand \@@endlink[0]{}%
\providecommand \url  [0]{\begingroup\@sanitize@url \@url }%
\providecommand \@url [1]{\endgroup\@href {#1}{\urlprefix }}%
\providecommand \urlprefix  [0]{URL }%
\providecommand \Eprint [0]{\href }%
\providecommand \doibase [0]{https://doi.org/}%
\providecommand \selectlanguage [0]{\@gobble}%
\providecommand \bibinfo  [0]{\@secondoftwo}%
\providecommand \bibfield  [0]{\@secondoftwo}%
\providecommand \translation [1]{[#1]}%
\providecommand \BibitemOpen [0]{}%
\providecommand \bibitemStop [0]{}%
\providecommand \bibitemNoStop [0]{.\EOS\space}%
\providecommand \EOS [0]{\spacefactor3000\relax}%
\providecommand \BibitemShut  [1]{\csname bibitem#1\endcsname}%
\let\auto@bib@innerbib\@empty
%</preamble>
\bibitem [{\citenamefont {Abbott}\ \emph
  {et~al.}(2021{\natexlab{a}})\citenamefont {Abbott} \emph
  {et~al.}}]{LIGOScientific:2021djp}%
  \BibitemOpen
  \bibfield  {author} {\bibinfo {author} {\bibfnamefont {R.}~\bibnamefont
  {Abbott}} \emph {et~al.} (\bibinfo {collaboration} {LIGO Scientific, VIRGO,
  KAGRA}),\ }\href@noop {} {\bibfield  {journal} {\bibinfo  {journal} {arXiv
  eprint}\ } (\bibinfo {year} {2021}{\natexlab{a}})},\ \Eprint
  {https://arxiv.org/abs/2111.03606} {arXiv:2111.03606 [gr-qc]} \BibitemShut
  {NoStop}%
\bibitem [{\citenamefont {Aasi}\ \emph {et~al.}(2015)\citenamefont {Aasi} \emph
  {et~al.}}]{LIGO}%
  \BibitemOpen
  \bibfield  {author} {\bibinfo {author} {\bibfnamefont {J.}~\bibnamefont
  {Aasi}} \emph {et~al.} (\bibinfo {collaboration} {LIGO Scientific
  Collaboration}),\ }\href {https://doi.org/10.1088/0264-9381/32/7/074001}
  {\bibfield  {journal} {\bibinfo  {journal} {Class. Quant. Grav.}\ }\textbf
  {\bibinfo {volume} {32}},\ \bibinfo {pages} {074001} (\bibinfo {year}
  {2015})},\ \Eprint {https://arxiv.org/abs/1411.4547} {arXiv:1411.4547
  [gr-qc]} \BibitemShut {NoStop}%
%%CITATION = ARXIV:1411.4547;%%
\bibitem [{\citenamefont {Acernese}\ \emph {et~al.}(2015)\citenamefont
  {Acernese} \emph {et~al.}}]{Virgo}%
  \BibitemOpen
  \bibfield  {author} {\bibinfo {author} {\bibfnamefont {F.}~\bibnamefont
  {Acernese}} \emph {et~al.} (\bibinfo {collaboration} {Virgo Collaboration}),\
  }\href {https://doi.org/10.1088/0264-9381/32/2/024001} {\bibfield  {journal}
  {\bibinfo  {journal} {Class. Quant. Grav.}\ }\textbf {\bibinfo {volume}
  {32}},\ \bibinfo {pages} {024001} (\bibinfo {year} {2015})},\ \Eprint
  {https://arxiv.org/abs/1408.3978} {arXiv:1408.3978 [gr-qc]} \BibitemShut
  {NoStop}%
%%CITATION = ARXIV:1408.3978;%%
\bibitem [{\citenamefont {Allen}\ and\ \citenamefont
  {Romano}(1999)}]{Allen:1997ad}%
  \BibitemOpen
  \bibfield  {author} {\bibinfo {author} {\bibfnamefont {B.}~\bibnamefont
  {Allen}}and\ \bibinfo {author} {\bibfnamefont {J.~D.}\ \bibnamefont
  {Romano}},\ }\href {https://doi.org/10.1103/PhysRevD.59.102001} {\bibfield
  {journal} {\bibinfo  {journal} {Phys. Rev. D}\ }\textbf {\bibinfo {volume}
  {59}},\ \bibinfo {pages} {102001} (\bibinfo {year} {1999})},\ \Eprint
  {https://arxiv.org/abs/gr-qc/9710117} {arXiv:gr-qc/9710117} \BibitemShut
  {NoStop}%
\bibitem [{\citenamefont {Romano}\ and\ \citenamefont
  {Cornish}(2017)}]{Romano:2016dpx}%
  \BibitemOpen
  \bibfield  {author} {\bibinfo {author} {\bibfnamefont {J.~D.}\ \bibnamefont
  {Romano}}and\ \bibinfo {author} {\bibfnamefont {N.~J.}\ \bibnamefont
  {Cornish}},\ }\href {https://doi.org/10.1007/s41114-017-0004-1} {\bibfield
  {journal} {\bibinfo  {journal} {Living Rev. Rel.}\ }\textbf {\bibinfo
  {volume} {20}},\ \bibinfo {pages} {2} (\bibinfo {year} {2017})},\ \Eprint
  {https://arxiv.org/abs/1608.06889} {arXiv:1608.06889 [gr-qc]} \BibitemShut
  {NoStop}%
\bibitem [{\citenamefont {Abbott}\ \emph
  {et~al.}(2021{\natexlab{b}})\citenamefont {Abbott} \emph
  {et~al.}}]{KAGRA:2021kbb}%
  \BibitemOpen
  \bibfield  {author} {\bibinfo {author} {\bibfnamefont {R.}~\bibnamefont
  {Abbott}} \emph {et~al.} (\bibinfo {collaboration} {KAGRA, Virgo, LIGO
  Scientific}),\ }\href {https://doi.org/10.1103/PhysRevD.104.022004}
  {\bibfield  {journal} {\bibinfo  {journal} {Phys. Rev. D}\ }\textbf {\bibinfo
  {volume} {104}},\ \bibinfo {pages} {022004} (\bibinfo {year}
  {2021}{\natexlab{b}})},\ \Eprint {https://arxiv.org/abs/2101.12130}
  {arXiv:2101.12130 [gr-qc]} \BibitemShut {NoStop}%
\bibitem [{\citenamefont {Abbott}\ \emph {et~al.}(2023)\citenamefont {Abbott},
  \citenamefont {Abbott}, \citenamefont {Acernese}, \citenamefont {Ackley},
  \citenamefont {Adams}, \citenamefont {Adhikari}, \citenamefont {Adhikari},
  \citenamefont {Adya}, \citenamefont {Affeldt} \emph
  {et~al.}}]{PhysRevX.13.011048}%
  \BibitemOpen
  \bibfield  {author} {\bibinfo {author} {\bibfnamefont {R.}~\bibnamefont
  {Abbott}}, \bibinfo {author} {\bibfnamefont {T.~D.}\ \bibnamefont {Abbott}},
  \bibinfo {author} {\bibfnamefont {F.}~\bibnamefont {Acernese}}, \bibinfo
  {author} {\bibfnamefont {K.}~\bibnamefont {Ackley}}, \bibinfo {author}
  {\bibfnamefont {C.}~\bibnamefont {Adams}}, \bibinfo {author} {\bibfnamefont
  {N.}~\bibnamefont {Adhikari}}, \bibinfo {author} {\bibfnamefont {R.~X.}\
  \bibnamefont {Adhikari}}, \bibinfo {author} {\bibfnamefont {V.~B.}\
  \bibnamefont {Adya}}, \bibinfo {author} {\bibfnamefont {C.}~\bibnamefont
  {Affeldt}},  \emph {et~al.} (\bibinfo {collaboration} {LIGO Scientific
  Collaboration, Virgo Collaboration, and KAGRA Collaboration}),\ }\href
  {https://doi.org/10.1103/PhysRevX.13.011048} {\bibfield  {journal} {\bibinfo
  {journal} {Phys. Rev. X}\ }\textbf {\bibinfo {volume} {13}},\ \bibinfo
  {pages} {011048} (\bibinfo {year} {2023})}\BibitemShut {NoStop}%
\bibitem [{\citenamefont {Renzini}\ \emph {et~al.}(2022)\citenamefont
  {Renzini}, \citenamefont {Goncharov}, \citenamefont {Jenkins},\ and\
  \citenamefont {Meyers}}]{Renzini:2022alw}%
  \BibitemOpen
  \bibfield  {author} {\bibinfo {author} {\bibfnamefont {A.~I.}\ \bibnamefont
  {Renzini}}, \bibinfo {author} {\bibfnamefont {B.}~\bibnamefont {Goncharov}},
  \bibinfo {author} {\bibfnamefont {A.~C.}\ \bibnamefont {Jenkins}}, and\
  \bibinfo {author} {\bibfnamefont {P.~M.}\ \bibnamefont {Meyers}},\ }\href
  {https://doi.org/10.3390/galaxies10010034} {\bibfield  {journal} {\bibinfo
  {journal} {Galaxies}\ }\textbf {\bibinfo {volume} {10}},\ \bibinfo {pages}
  {34} (\bibinfo {year} {2022})},\ \Eprint {https://arxiv.org/abs/2202.00178}
  {arXiv:2202.00178 [gr-qc]} \BibitemShut {NoStop}%
\bibitem [{\citenamefont {Callister}\ \emph {et~al.}(2016)\citenamefont
  {Callister}, \citenamefont {Sammut}, \citenamefont {Qiu}, \citenamefont
  {Mandel},\ and\ \citenamefont {Thrane}}]{Callister:2016ewt}%
  \BibitemOpen
  \bibfield  {author} {\bibinfo {author} {\bibfnamefont {T.}~\bibnamefont
  {Callister}}, \bibinfo {author} {\bibfnamefont {L.}~\bibnamefont {Sammut}},
  \bibinfo {author} {\bibfnamefont {S.}~\bibnamefont {Qiu}}, \bibinfo {author}
  {\bibfnamefont {I.}~\bibnamefont {Mandel}}, and\ \bibinfo {author}
  {\bibfnamefont {E.}~\bibnamefont {Thrane}},\ }\href
  {https://doi.org/10.1103/PhysRevX.6.031018} {\bibfield  {journal} {\bibinfo
  {journal} {Phys. Rev. X}\ }\textbf {\bibinfo {volume} {6}},\ \bibinfo {pages}
  {031018} (\bibinfo {year} {2016})},\ \Eprint
  {https://arxiv.org/abs/1604.02513} {arXiv:1604.02513 [gr-qc]} \BibitemShut
  {NoStop}%
\bibitem [{\citenamefont {Callister}\ \emph {et~al.}(2020)\citenamefont
  {Callister}, \citenamefont {Fishbach}, \citenamefont {Holz},\ and\
  \citenamefont {Farr}}]{Callister:2020arv}%
  \BibitemOpen
  \bibfield  {author} {\bibinfo {author} {\bibfnamefont {T.}~\bibnamefont
  {Callister}}, \bibinfo {author} {\bibfnamefont {M.}~\bibnamefont {Fishbach}},
  \bibinfo {author} {\bibfnamefont {D.}~\bibnamefont {Holz}}, and\ \bibinfo
  {author} {\bibfnamefont {W.}~\bibnamefont {Farr}},\ }\href
  {https://doi.org/10.3847/2041-8213/ab9743} {\bibfield  {journal} {\bibinfo
  {journal} {Astrophys. J. Lett.}\ }\textbf {\bibinfo {volume} {896}},\
  \bibinfo {pages} {L32} (\bibinfo {year} {2020})},\ \Eprint
  {https://arxiv.org/abs/2003.12152} {arXiv:2003.12152 [astro-ph.HE]}
  \BibitemShut {NoStop}%
\bibitem [{\citenamefont {Bavera}\ \emph {et~al.}(2022)\citenamefont {Bavera},
  \citenamefont {Franciolini}, \citenamefont {Cusin}, \citenamefont {Riotto},
  \citenamefont {Zevin},\ and\ \citenamefont {Fragos}}]{Bavera:2021wmw}%
  \BibitemOpen
  \bibfield  {author} {\bibinfo {author} {\bibfnamefont {S.~S.}\ \bibnamefont
  {Bavera}}, \bibinfo {author} {\bibfnamefont {G.}~\bibnamefont {Franciolini}},
  \bibinfo {author} {\bibfnamefont {G.}~\bibnamefont {Cusin}}, \bibinfo
  {author} {\bibfnamefont {A.}~\bibnamefont {Riotto}}, \bibinfo {author}
  {\bibfnamefont {M.}~\bibnamefont {Zevin}}, and\ \bibinfo {author}
  {\bibfnamefont {T.}~\bibnamefont {Fragos}},\ }\href
  {https://doi.org/10.1051/0004-6361/202142208} {\bibfield  {journal} {\bibinfo
   {journal} {Astron. Astrophys.}\ }\textbf {\bibinfo {volume} {660}},\
  \bibinfo {pages} {A26} (\bibinfo {year} {2022})},\ \Eprint
  {https://arxiv.org/abs/2109.05836} {arXiv:2109.05836 [astro-ph.CO]}
  \BibitemShut {NoStop}%
\bibitem [{\citenamefont {Lehoucq}\ \emph {et~al.}(2023)\citenamefont
  {Lehoucq}, \citenamefont {Dvorkin}, \citenamefont {Srinivasan}, \citenamefont
  {Pellouin},\ and\ \citenamefont {Lamberts}}]{Lehoucq:2023zlt}%
  \BibitemOpen
  \bibfield  {author} {\bibinfo {author} {\bibfnamefont {L.}~\bibnamefont
  {Lehoucq}}, \bibinfo {author} {\bibfnamefont {I.}~\bibnamefont {Dvorkin}},
  \bibinfo {author} {\bibfnamefont {R.}~\bibnamefont {Srinivasan}}, \bibinfo
  {author} {\bibfnamefont {C.}~\bibnamefont {Pellouin}}, and\ \bibinfo {author}
  {\bibfnamefont {A.}~\bibnamefont {Lamberts}},\ }\href
  {https://doi.org/10.1093/mnras/stad2917} {\bibfield  {journal} {\bibinfo
  {journal} {Mon. Not. Roy. Astron. Soc.}\ }\textbf {\bibinfo {volume} {526}},\
  \bibinfo {pages} {4378} (\bibinfo {year} {2023})},\ \Eprint
  {https://arxiv.org/abs/2306.09861} {arXiv:2306.09861 [astro-ph.HE]}
  \BibitemShut {NoStop}%
\bibitem [{\citenamefont {Turbang}\ \emph {et~al.}(2023)\citenamefont
  {Turbang}, \citenamefont {Lalleman}, \citenamefont {Callister},\ and\
  \citenamefont {van Remortel}}]{Turbang:2023tjk}%
  \BibitemOpen
  \bibfield  {author} {\bibinfo {author} {\bibfnamefont {K.}~\bibnamefont
  {Turbang}}, \bibinfo {author} {\bibfnamefont {M.}~\bibnamefont {Lalleman}},
  \bibinfo {author} {\bibfnamefont {T.~A.}\ \bibnamefont {Callister}}, and\
  \bibinfo {author} {\bibfnamefont {N.}~\bibnamefont {van Remortel}},\
  }\href@noop {} {\bibfield  {journal} {\bibinfo  {journal} {arXiv eprints}\ }
  (\bibinfo {year} {2023})},\ \Eprint {https://arxiv.org/abs/2310.17625}
  {arXiv:2310.17625 [astro-ph.HE]} \BibitemShut {NoStop}%
\bibitem [{\citenamefont {Abbott}\ \emph
  {et~al.}(2019{\natexlab{a}})\citenamefont {Abbott} \emph
  {et~al.}}]{LIGOScientific:2019vic}%
  \BibitemOpen
  \bibfield  {author} {\bibinfo {author} {\bibfnamefont {B.~P.}\ \bibnamefont
  {Abbott}} \emph {et~al.} (\bibinfo {collaboration} {LIGO Scientific,
  Virgo}),\ }\href {https://doi.org/10.1103/PhysRevD.100.061101} {\bibfield
  {journal} {\bibinfo  {journal} {Phys. Rev. D}\ }\textbf {\bibinfo {volume}
  {100}},\ \bibinfo {pages} {061101} (\bibinfo {year} {2019}{\natexlab{a}})},\
  \Eprint {https://arxiv.org/abs/1903.02886} {arXiv:1903.02886 [gr-qc]}
  \BibitemShut {NoStop}%
\bibitem [{\citenamefont {Abbott}\ \emph
  {et~al.}(2021{\natexlab{c}})\citenamefont {Abbott} \emph
  {et~al.}}]{KAGRA:2021mth}%
  \BibitemOpen
  \bibfield  {author} {\bibinfo {author} {\bibfnamefont {R.}~\bibnamefont
  {Abbott}} \emph {et~al.} (\bibinfo {collaboration} {KAGRA, Virgo, LIGO
  Scientific}),\ }\href {https://doi.org/10.1103/PhysRevD.104.022005}
  {\bibfield  {journal} {\bibinfo  {journal} {Phys. Rev. D}\ }\textbf {\bibinfo
  {volume} {104}},\ \bibinfo {pages} {022005} (\bibinfo {year}
  {2021}{\natexlab{c}})},\ \Eprint {https://arxiv.org/abs/2103.08520}
  {arXiv:2103.08520 [gr-qc]} \BibitemShut {NoStop}%
\bibitem [{\citenamefont {Buikema}\ \emph {et~al.}(2020)\citenamefont {Buikema}
  \emph {et~al.}}]{aLIGO:2020wna}%
  \BibitemOpen
  \bibfield  {author} {\bibinfo {author} {\bibfnamefont {A.}~\bibnamefont
  {Buikema}} \emph {et~al.} (\bibinfo {collaboration} {aLIGO}),\ }\href
  {https://doi.org/10.1103/PhysRevD.102.062003} {\bibfield  {journal} {\bibinfo
   {journal} {Phys. Rev. D}\ }\textbf {\bibinfo {volume} {102}},\ \bibinfo
  {pages} {062003} (\bibinfo {year} {2020})},\ \Eprint
  {https://arxiv.org/abs/2008.01301} {arXiv:2008.01301 [astro-ph.IM]}
  \BibitemShut {NoStop}%
\bibitem [{\citenamefont {Matas}\ and\ \citenamefont
  {Romano}(2021)}]{Matas:2020roi}%
  \BibitemOpen
  \bibfield  {author} {\bibinfo {author} {\bibfnamefont {A.}~\bibnamefont
  {Matas}}and\ \bibinfo {author} {\bibfnamefont {J.~D.}\ \bibnamefont
  {Romano}},\ }\href {https://doi.org/10.1103/PhysRevD.103.062003} {\bibfield
  {journal} {\bibinfo  {journal} {Phys. Rev. D}\ }\textbf {\bibinfo {volume}
  {103}},\ \bibinfo {pages} {062003} (\bibinfo {year} {2021})},\ \Eprint
  {https://arxiv.org/abs/2012.00907} {arXiv:2012.00907 [gr-qc]} \BibitemShut
  {NoStop}%
\bibitem [{\citenamefont {Abbott}\ \emph
  {et~al.}(2018{\natexlab{a}})\citenamefont {Abbott} \emph
  {et~al.}}]{KAGRA:2013rdx}%
  \BibitemOpen
  \bibfield  {author} {\bibinfo {author} {\bibfnamefont {B.~P.}\ \bibnamefont
  {Abbott}} \emph {et~al.} (\bibinfo {collaboration} {KAGRA, LIGO Scientific,
  Virgo, VIRGO}),\ }\href {https://doi.org/10.1007/s41114-020-00026-9}
  {\bibfield  {journal} {\bibinfo  {journal} {Living Rev. Rel.}\ }\textbf
  {\bibinfo {volume} {21}},\ \bibinfo {pages} {3} (\bibinfo {year}
  {2018}{\natexlab{a}})},\ \Eprint {https://arxiv.org/abs/1304.0670}
  {arXiv:1304.0670 [gr-qc]} \BibitemShut {NoStop}%
\bibitem [{\citenamefont {Abbott}\ \emph
  {et~al.}(2019{\natexlab{b}})\citenamefont {Abbott} \emph
  {et~al.}}]{LIGOScientific:2018mvr}%
  \BibitemOpen
  \bibfield  {author} {\bibinfo {author} {\bibfnamefont {B.~P.}\ \bibnamefont
  {Abbott}} \emph {et~al.} (\bibinfo {collaboration} {LIGO Scientific,
  Virgo}),\ }\href {https://doi.org/10.1103/PhysRevX.9.031040} {\bibfield
  {journal} {\bibinfo  {journal} {Phys. Rev. X}\ }\textbf {\bibinfo {volume}
  {9}},\ \bibinfo {pages} {031040} (\bibinfo {year} {2019}{\natexlab{b}})},\
  \Eprint {https://arxiv.org/abs/1811.12907} {arXiv:1811.12907 [astro-ph.HE]}
  \BibitemShut {NoStop}%
\bibitem [{\citenamefont {Abbott}\ \emph
  {et~al.}(2021{\natexlab{d}})\citenamefont {Abbott} \emph
  {et~al.}}]{LIGOScientific:2020ibl}%
  \BibitemOpen
  \bibfield  {author} {\bibinfo {author} {\bibfnamefont {R.}~\bibnamefont
  {Abbott}} \emph {et~al.} (\bibinfo {collaboration} {LIGO Scientific,
  Virgo}),\ }\href {https://doi.org/10.1103/PhysRevX.11.021053} {\bibfield
  {journal} {\bibinfo  {journal} {Phys. Rev. X}\ }\textbf {\bibinfo {volume}
  {11}},\ \bibinfo {pages} {021053} (\bibinfo {year} {2021}{\natexlab{d}})},\
  \Eprint {https://arxiv.org/abs/2010.14527} {arXiv:2010.14527 [gr-qc]}
  \BibitemShut {NoStop}%
\bibitem [{\citenamefont {Abbott}\ \emph
  {et~al.}(2019{\natexlab{c}})\citenamefont {Abbott} \emph
  {et~al.}}]{LIGOScientific:2018hze}%
  \BibitemOpen
  \bibfield  {author} {\bibinfo {author} {\bibfnamefont {B.~P.}\ \bibnamefont
  {Abbott}} \emph {et~al.} (\bibinfo {collaboration} {LIGO Scientific,
  Virgo}),\ }\href {https://doi.org/10.1103/PhysRevX.9.011001} {\bibfield
  {journal} {\bibinfo  {journal} {Phys. Rev. X}\ }\textbf {\bibinfo {volume}
  {9}},\ \bibinfo {pages} {011001} (\bibinfo {year} {2019}{\natexlab{c}})},\
  \Eprint {https://arxiv.org/abs/1805.11579} {arXiv:1805.11579 [gr-qc]}
  \BibitemShut {NoStop}%
\bibitem [{\citenamefont {Regimbau}(2007)}]{Regimbau:2007cw}%
  \BibitemOpen
  \bibfield  {author} {\bibinfo {author} {\bibfnamefont {T.}~\bibnamefont
  {Regimbau}},\ }\href {https://doi.org/10.1103/PhysRevD.75.043002} {\bibfield
  {journal} {\bibinfo  {journal} {Phys. Rev. D}\ }\textbf {\bibinfo {volume}
  {75}},\ \bibinfo {pages} {043002} (\bibinfo {year} {2007})},\ \Eprint
  {https://arxiv.org/abs/astro-ph/0701004} {arXiv:astro-ph/0701004}
  \BibitemShut {NoStop}%
\bibitem [{\citenamefont {Meacher}\ \emph {et~al.}(2014)\citenamefont
  {Meacher}, \citenamefont {Thrane},\ and\ \citenamefont
  {Regimbau}}]{Meacher:2014aca}%
  \BibitemOpen
  \bibfield  {author} {\bibinfo {author} {\bibfnamefont {D.}~\bibnamefont
  {Meacher}}, \bibinfo {author} {\bibfnamefont {E.}~\bibnamefont {Thrane}},
  and\ \bibinfo {author} {\bibfnamefont {T.}~\bibnamefont {Regimbau}},\ }\href
  {https://doi.org/10.1103/PhysRevD.89.084063} {\bibfield  {journal} {\bibinfo
  {journal} {Phys. Rev. D}\ }\textbf {\bibinfo {volume} {89}},\ \bibinfo
  {pages} {084063} (\bibinfo {year} {2014})},\ \Eprint
  {https://arxiv.org/abs/1402.6231} {arXiv:1402.6231 [astro-ph.CO]}
  \BibitemShut {NoStop}%
\bibitem [{\citenamefont {Abbott}\ \emph
  {et~al.}(2018{\natexlab{b}})\citenamefont {Abbott} \emph
  {et~al.}}]{LIGOScientific:2017zlf}%
  \BibitemOpen
  \bibfield  {author} {\bibinfo {author} {\bibfnamefont {B.~P.}\ \bibnamefont
  {Abbott}} \emph {et~al.} (\bibinfo {collaboration} {LIGO Scientific,
  Virgo}),\ }\href {https://doi.org/10.1103/PhysRevLett.120.091101} {\bibfield
  {journal} {\bibinfo  {journal} {Phys. Rev. Lett.}\ }\textbf {\bibinfo
  {volume} {120}},\ \bibinfo {pages} {091101} (\bibinfo {year}
  {2018}{\natexlab{b}})},\ \Eprint {https://arxiv.org/abs/1710.05837}
  {arXiv:1710.05837 [gr-qc]} \BibitemShut {NoStop}%
\bibitem [{\citenamefont {Zhong}\ \emph {et~al.}(2023)\citenamefont {Zhong},
  \citenamefont {Ormiston},\ and\ \citenamefont {Mandic}}]{Zhong:2022ylh}%
  \BibitemOpen
  \bibfield  {author} {\bibinfo {author} {\bibfnamefont {H.}~\bibnamefont
  {Zhong}}, \bibinfo {author} {\bibfnamefont {R.}~\bibnamefont {Ormiston}},
  and\ \bibinfo {author} {\bibfnamefont {V.}~\bibnamefont {Mandic}},\ }\href
  {https://doi.org/10.1103/PhysRevD.107.064048} {\bibfield  {journal} {\bibinfo
   {journal} {Phys. Rev. D}\ }\textbf {\bibinfo {volume} {107}},\ \bibinfo
  {pages} {064048} (\bibinfo {year} {2023})},\ \Eprint
  {https://arxiv.org/abs/2209.11877} {arXiv:2209.11877 [gr-qc]} \BibitemShut
  {NoStop}%
\bibitem [{\citenamefont {Johnson}\ \emph {et~al.}(2024)\citenamefont
  {Johnson}, \citenamefont {Chatziioannou},\ and\ \citenamefont
  {Farr}}]{Johnson:2024foj}%
  \BibitemOpen
  \bibfield  {author} {\bibinfo {author} {\bibfnamefont {A.~D.}\ \bibnamefont
  {Johnson}}, \bibinfo {author} {\bibfnamefont {K.}~\bibnamefont
  {Chatziioannou}}, and\ \bibinfo {author} {\bibfnamefont {W.~M.}\ \bibnamefont
  {Farr}},\ }\href@noop {} {\bibfield  {journal} {\bibinfo  {journal} {arXiv
  eprints}\ } (\bibinfo {year} {2024})},\ \Eprint
  {https://arxiv.org/abs/2402.06836} {arXiv:2402.06836 [gr-qc]} \BibitemShut
  {NoStop}%
\bibitem [{\citenamefont {Drasco}\ and\ \citenamefont
  {Flanagan}(2003)}]{Drasco:2002yd}%
  \BibitemOpen
  \bibfield  {author} {\bibinfo {author} {\bibfnamefont {S.}~\bibnamefont
  {Drasco}}and\ \bibinfo {author} {\bibfnamefont {E.~E.}\ \bibnamefont
  {Flanagan}},\ }\href {https://doi.org/10.1103/PhysRevD.67.082003} {\bibfield
  {journal} {\bibinfo  {journal} {Phys. Rev. D}\ }\textbf {\bibinfo {volume}
  {67}},\ \bibinfo {pages} {082003} (\bibinfo {year} {2003})},\ \Eprint
  {https://arxiv.org/abs/gr-qc/0210032} {arXiv:gr-qc/0210032} \BibitemShut
  {NoStop}%
\bibitem [{\citenamefont {Smith}\ and\ \citenamefont
  {Thrane}(2018)}]{Smith:2017vfk}%
  \BibitemOpen
  \bibfield  {author} {\bibinfo {author} {\bibfnamefont {R.}~\bibnamefont
  {Smith}}and\ \bibinfo {author} {\bibfnamefont {E.}~\bibnamefont {Thrane}},\
  }\href {https://doi.org/10.1103/PhysRevX.8.021019} {\bibfield  {journal}
  {\bibinfo  {journal} {Phys. Rev. X}\ }\textbf {\bibinfo {volume} {8}},\
  \bibinfo {pages} {021019} (\bibinfo {year} {2018})},\ \Eprint
  {https://arxiv.org/abs/1712.00688} {arXiv:1712.00688 [gr-qc]} \BibitemShut
  {NoStop}%
\bibitem [{\citenamefont {Abbott}\ \emph {et~al.}(2016)\citenamefont {Abbott}
  \emph {et~al.}}]{LIGOScientific:2016vbw}%
  \BibitemOpen
  \bibfield  {author} {\bibinfo {author} {\bibfnamefont {B.~P.}\ \bibnamefont
  {Abbott}} \emph {et~al.} (\bibinfo {collaboration} {LIGO Scientific,
  Virgo}),\ }\href {https://doi.org/10.1103/PhysRevD.93.122003} {\bibfield
  {journal} {\bibinfo  {journal} {Phys. Rev. D}\ }\textbf {\bibinfo {volume}
  {93}},\ \bibinfo {pages} {122003} (\bibinfo {year} {2016})},\ \Eprint
  {https://arxiv.org/abs/1602.03839} {arXiv:1602.03839 [gr-qc]} \BibitemShut
  {NoStop}%
\bibitem [{\citenamefont {Smith}\ \emph {et~al.}(2020)\citenamefont {Smith},
  \citenamefont {Talbot}, \citenamefont {Hernandez~Vivanco},\ and\
  \citenamefont {Thrane}}]{Smith:2020lkj}%
  \BibitemOpen
  \bibfield  {author} {\bibinfo {author} {\bibfnamefont {R.~J.~E.}\
  \bibnamefont {Smith}}, \bibinfo {author} {\bibfnamefont {C.}~\bibnamefont
  {Talbot}}, \bibinfo {author} {\bibfnamefont {F.}~\bibnamefont
  {Hernandez~Vivanco}}, and\ \bibinfo {author} {\bibfnamefont {E.}~\bibnamefont
  {Thrane}},\ }\href {https://doi.org/10.1093/mnras/staa1642} {\bibfield
  {journal} {\bibinfo  {journal} {Mon. Not. Roy. Astron. Soc.}\ }\textbf
  {\bibinfo {volume} {496}},\ \bibinfo {pages} {3281} (\bibinfo {year}
  {2020})},\ \Eprint {https://arxiv.org/abs/2004.09700} {arXiv:2004.09700
  [astro-ph.HE]} \BibitemShut {NoStop}%
\bibitem [{\citenamefont {Regimbau}(2011)}]{Regimbau:2011rp}%
  \BibitemOpen
  \bibfield  {author} {\bibinfo {author} {\bibfnamefont {T.}~\bibnamefont
  {Regimbau}},\ }\href {https://doi.org/10.1088/1674-4527/11/4/001} {\bibfield
  {journal} {\bibinfo  {journal} {Res. Astron. Astrophys.}\ }\textbf {\bibinfo
  {volume} {11}},\ \bibinfo {pages} {369} (\bibinfo {year} {2011})},\ \Eprint
  {https://arxiv.org/abs/1101.2762} {arXiv:1101.2762 [astro-ph.CO]}
  \BibitemShut {NoStop}%
\bibitem [{\citenamefont {{Christensen}}(1992)}]{1992PhRvD..46.5250C}%
  \BibitemOpen
  \bibfield  {author} {\bibinfo {author} {\bibfnamefont {N.}~\bibnamefont
  {{Christensen}}},\ }\href {https://doi.org/10.1103/PhysRevD.46.5250}
  {\bibfield  {journal} {\bibinfo  {journal} {\prd}\ }\textbf {\bibinfo
  {volume} {46}},\ \bibinfo {pages} {5250} (\bibinfo {year}
  {1992})}\BibitemShut {NoStop}%
\bibitem [{\citenamefont {Flanagan}(1993)}]{flanagan_sensitivity_1993}%
  \BibitemOpen
  \bibfield  {author} {\bibinfo {author} {\bibfnamefont {E.~E.}\ \bibnamefont
  {Flanagan}},\ }\href {https://doi.org/10.1103/PhysRevD.48.2389} {\bibfield
  {journal} {\bibinfo  {journal} {Physical Review D}\ }\textbf {\bibinfo
  {volume} {48}},\ \bibinfo {pages} {2389} (\bibinfo {year}
  {1993})}\BibitemShut {NoStop}%
\bibitem [{\citenamefont {Mandic}\ \emph {et~al.}(2012)\citenamefont {Mandic},
  \citenamefont {Thrane}, \citenamefont {Giampanis},\ and\ \citenamefont
  {Regimbau}}]{Mandic:2012pj}%
  \BibitemOpen
  \bibfield  {author} {\bibinfo {author} {\bibfnamefont {V.}~\bibnamefont
  {Mandic}}, \bibinfo {author} {\bibfnamefont {E.}~\bibnamefont {Thrane}},
  \bibinfo {author} {\bibfnamefont {S.}~\bibnamefont {Giampanis}}, and\
  \bibinfo {author} {\bibfnamefont {T.}~\bibnamefont {Regimbau}},\ }\href
  {https://doi.org/10.1103/PhysRevLett.109.171102} {\bibfield  {journal}
  {\bibinfo  {journal} {Phys. Rev. Lett.}\ }\textbf {\bibinfo {volume} {109}},\
  \bibinfo {pages} {171102} (\bibinfo {year} {2012})},\ \Eprint
  {https://arxiv.org/abs/1209.3847} {arXiv:1209.3847 [astro-ph.CO]}
  \BibitemShut {NoStop}%
\bibitem [{\citenamefont {Hernandez~Vivanco}\ \emph {et~al.}(2019)\citenamefont
  {Hernandez~Vivanco}, \citenamefont {Smith}, \citenamefont {Thrane},\ and\
  \citenamefont {Lasky}}]{HernandezVivanco:2019fku}%
  \BibitemOpen
  \bibfield  {author} {\bibinfo {author} {\bibfnamefont {F.}~\bibnamefont
  {Hernandez~Vivanco}}, \bibinfo {author} {\bibfnamefont {R.}~\bibnamefont
  {Smith}}, \bibinfo {author} {\bibfnamefont {E.}~\bibnamefont {Thrane}}, and\
  \bibinfo {author} {\bibfnamefont {P.~D.}\ \bibnamefont {Lasky}},\ }\href
  {https://doi.org/10.1103/PhysRevD.100.043023} {\bibfield  {journal} {\bibinfo
   {journal} {Phys. Rev. D}\ }\textbf {\bibinfo {volume} {100}},\ \bibinfo
  {pages} {043023} (\bibinfo {year} {2019})},\ \Eprint
  {https://arxiv.org/abs/1903.05778} {arXiv:1903.05778 [gr-qc]} \BibitemShut
  {NoStop}%
\bibitem [{\citenamefont {Renzini}\ \emph {et~al.}(2023)\citenamefont {Renzini}
  \emph {et~al.}}]{Renzini:2023qtj}%
  \BibitemOpen
  \bibfield  {author} {\bibinfo {author} {\bibfnamefont {A.~I.}\ \bibnamefont
  {Renzini}} \emph {et~al.},\ }\href@noop {} {\bibfield  {journal} {\bibinfo
  {journal} {arXiv eprint}\ } (\bibinfo {year} {2023})},\ \Eprint
  {https://arxiv.org/abs/2303.15696} {arXiv:2303.15696 [gr-qc]} \BibitemShut
  {NoStop}%
\bibitem [{\citenamefont {Abbott}\ \emph
  {et~al.}(2021{\natexlab{e}})\citenamefont {Abbott} \emph
  {et~al.}}]{LIGOScientific:2020kqk}%
  \BibitemOpen
  \bibfield  {author} {\bibinfo {author} {\bibfnamefont {R.}~\bibnamefont
  {Abbott}} \emph {et~al.} (\bibinfo {collaboration} {LIGO Scientific,
  Virgo}),\ }\href {https://doi.org/10.3847/2041-8213/abe949} {\bibfield
  {journal} {\bibinfo  {journal} {Astrophys. J. Lett.}\ }\textbf {\bibinfo
  {volume} {913}},\ \bibinfo {pages} {L7} (\bibinfo {year}
  {2021}{\natexlab{e}})},\ \Eprint {https://arxiv.org/abs/2010.14533}
  {arXiv:2010.14533 [astro-ph.HE]} \BibitemShut {NoStop}%
\bibitem [{des(2019)}]{design_sensitivity_curve}%
  \BibitemOpen
  \href@noop {} {}\bibinfo {howpublished}
  {\url{https://dcc.ligo.org/LIGO-P1200087/public}} (\bibinfo {year}
  {2019})\BibitemShut {NoStop}%
\bibitem [{\citenamefont {Zhu}\ \emph {et~al.}(2011)\citenamefont {Zhu},
  \citenamefont {Howell}, \citenamefont {Regimbau}, \citenamefont {Blair},\
  and\ \citenamefont {Zhu}}]{Zhu_2011}%
  \BibitemOpen
  \bibfield  {author} {\bibinfo {author} {\bibfnamefont {X.-J.}\ \bibnamefont
  {Zhu}}, \bibinfo {author} {\bibfnamefont {E.}~\bibnamefont {Howell}},
  \bibinfo {author} {\bibfnamefont {T.}~\bibnamefont {Regimbau}}, \bibinfo
  {author} {\bibfnamefont {D.}~\bibnamefont {Blair}}, and\ \bibinfo {author}
  {\bibfnamefont {Z.-H.}\ \bibnamefont {Zhu}},\ }\href
  {https://doi.org/10.1088/0004-637X/739/2/86} {\bibfield  {journal} {\bibinfo
  {journal} {The Astrophysical Journal}\ }\textbf {\bibinfo {volume} {739}},\
  \bibinfo {pages} {86} (\bibinfo {year} {2011})}\BibitemShut {NoStop}%
\bibitem [{\citenamefont {Ajith}\ \emph {et~al.}(2008)\citenamefont {Ajith}
  \emph {et~al.}}]{Ajith:2007kx}%
  \BibitemOpen
  \bibfield  {author} {\bibinfo {author} {\bibfnamefont {P.}~\bibnamefont
  {Ajith}} \emph {et~al.},\ }\href {https://doi.org/10.1103/PhysRevD.77.104017}
  {\bibfield  {journal} {\bibinfo  {journal} {Phys. Rev. D}\ }\textbf {\bibinfo
  {volume} {77}},\ \bibinfo {pages} {104017} (\bibinfo {year} {2008})},\
  \bibinfo {note} {[Erratum: Phys.Rev.D 79, 129901 (2009)]},\ \Eprint
  {https://arxiv.org/abs/0710.2335} {arXiv:0710.2335 [gr-qc]} \BibitemShut
  {NoStop}%
\bibitem [{\citenamefont {Ashton}\ \emph {et~al.}(2019)\citenamefont {Ashton}
  \emph {et~al.}}]{Ashton:2018jfp}%
  \BibitemOpen
  \bibfield  {author} {\bibinfo {author} {\bibfnamefont {G.}~\bibnamefont
  {Ashton}} \emph {et~al.},\ }\href {https://doi.org/10.3847/1538-4365/ab06fc}
  {\bibfield  {journal} {\bibinfo  {journal} {Astrophys. J. Suppl.}\ }\textbf
  {\bibinfo {volume} {241}},\ \bibinfo {pages} {27} (\bibinfo {year} {2019})},\
  \Eprint {https://arxiv.org/abs/1811.02042} {arXiv:1811.02042 [astro-ph.IM]}
  \BibitemShut {NoStop}%
\bibitem [{sen(2023)}]{sensitivity_curves}%
  \BibitemOpen
  \href@noop {} {}\bibinfo {howpublished}
  {\url{https://dcc.ligo.org/LIGO-T1800042/public}} (\bibinfo {year}
  {2023})\BibitemShut {NoStop}%
\bibitem [{\citenamefont {Lawrence}\ \emph {et~al.}(2023)\citenamefont
  {Lawrence}, \citenamefont {Turbang}, \citenamefont {Matas}, \citenamefont
  {Renzini}, \citenamefont {van Remortel},\ and\ \citenamefont
  {Romano}}]{Lawrence:2023buo}%
  \BibitemOpen
  \bibfield  {author} {\bibinfo {author} {\bibfnamefont {J.}~\bibnamefont
  {Lawrence}}, \bibinfo {author} {\bibfnamefont {K.}~\bibnamefont {Turbang}},
  \bibinfo {author} {\bibfnamefont {A.}~\bibnamefont {Matas}}, \bibinfo
  {author} {\bibfnamefont {A.~I.}\ \bibnamefont {Renzini}}, \bibinfo {author}
  {\bibfnamefont {N.}~\bibnamefont {van Remortel}}, and\ \bibinfo {author}
  {\bibfnamefont {J.~D.}\ \bibnamefont {Romano}},\ }\href
  {https://doi.org/10.1103/PhysRevD.107.103026} {\bibfield  {journal} {\bibinfo
   {journal} {Phys. Rev. D}\ }\textbf {\bibinfo {volume} {107}},\ \bibinfo
  {pages} {103026} (\bibinfo {year} {2023})},\ \Eprint
  {https://arxiv.org/abs/2301.07675} {arXiv:2301.07675 [gr-qc]} \BibitemShut
  {NoStop}%
\bibitem [{\citenamefont {Harris}\ \emph {et~al.}(2020)\citenamefont {Harris},
  \citenamefont {Millman}, \citenamefont {van~der Walt}, \citenamefont
  {Gommers}, \citenamefont {Virtanen}, \citenamefont {Cournapeau},
  \citenamefont {Wieser}, \citenamefont {Taylor}, \citenamefont {Berg},
  \citenamefont {Smith}, \citenamefont {Kern}, \citenamefont {Picus},
  \citenamefont {Hoyer}, \citenamefont {van Kerkwijk}, \citenamefont {Brett},
  \citenamefont {Haldane}, \citenamefont {del R{\'{i}}o}, \citenamefont
  {Wiebe}, \citenamefont {Peterson}, \citenamefont {G{\'{e}}rard-Marchant},
  \citenamefont {Sheppard}, \citenamefont {Reddy}, \citenamefont {Weckesser},
  \citenamefont {Abbasi}, \citenamefont {Gohlke},\ and\ \citenamefont
  {Oliphant}}]{harris2020array}%
  \BibitemOpen
  \bibfield  {author} {\bibinfo {author} {\bibfnamefont {C.~R.}\ \bibnamefont
  {Harris}}, \bibinfo {author} {\bibfnamefont {K.~J.}\ \bibnamefont {Millman}},
  \bibinfo {author} {\bibfnamefont {S.~J.}\ \bibnamefont {van~der Walt}},
  \bibinfo {author} {\bibfnamefont {R.}~\bibnamefont {Gommers}}, \bibinfo
  {author} {\bibfnamefont {P.}~\bibnamefont {Virtanen}}, \bibinfo {author}
  {\bibfnamefont {D.}~\bibnamefont {Cournapeau}}, \bibinfo {author}
  {\bibfnamefont {E.}~\bibnamefont {Wieser}}, \bibinfo {author} {\bibfnamefont
  {J.}~\bibnamefont {Taylor}}, \bibinfo {author} {\bibfnamefont
  {S.}~\bibnamefont {Berg}}, \bibinfo {author} {\bibfnamefont {N.~J.}\
  \bibnamefont {Smith}}, \bibinfo {author} {\bibfnamefont {R.}~\bibnamefont
  {Kern}}, \bibinfo {author} {\bibfnamefont {M.}~\bibnamefont {Picus}},
  \bibinfo {author} {\bibfnamefont {S.}~\bibnamefont {Hoyer}}, \bibinfo
  {author} {\bibfnamefont {M.~H.}\ \bibnamefont {van Kerkwijk}}, \bibinfo
  {author} {\bibfnamefont {M.}~\bibnamefont {Brett}}, \bibinfo {author}
  {\bibfnamefont {A.}~\bibnamefont {Haldane}}, \bibinfo {author} {\bibfnamefont
  {J.~F.}\ \bibnamefont {del R{\'{i}}o}}, \bibinfo {author} {\bibfnamefont
  {M.}~\bibnamefont {Wiebe}}, \bibinfo {author} {\bibfnamefont
  {P.}~\bibnamefont {Peterson}}, \bibinfo {author} {\bibfnamefont
  {P.}~\bibnamefont {G{\'{e}}rard-Marchant}}, \bibinfo {author} {\bibfnamefont
  {K.}~\bibnamefont {Sheppard}}, \bibinfo {author} {\bibfnamefont
  {T.}~\bibnamefont {Reddy}}, \bibinfo {author} {\bibfnamefont
  {W.}~\bibnamefont {Weckesser}}, \bibinfo {author} {\bibfnamefont
  {H.}~\bibnamefont {Abbasi}}, \bibinfo {author} {\bibfnamefont
  {C.}~\bibnamefont {Gohlke}}, and\ \bibinfo {author} {\bibfnamefont {T.~E.}\
  \bibnamefont {Oliphant}},\ }\href {https://doi.org/10.1038/s41586-020-2649-2}
  {\bibfield  {journal} {\bibinfo  {journal} {Nature}\ }\textbf {\bibinfo
  {volume} {585}},\ \bibinfo {pages} {357} (\bibinfo {year}
  {2020})}\BibitemShut {NoStop}%
\bibitem [{\citenamefont {Virtanen}\ \emph {et~al.}(2020)\citenamefont
  {Virtanen}, \citenamefont {Gommers}, \citenamefont {Oliphant}, \citenamefont
  {Haberland}, \citenamefont {Reddy}, \citenamefont {Cournapeau}, \citenamefont
  {Burovski}, \citenamefont {Peterson}, \citenamefont {Weckesser},
  \citenamefont {Bright}, \citenamefont {{van der Walt}}, \citenamefont
  {Brett}, \citenamefont {Wilson}, \citenamefont {Millman}, \citenamefont
  {Mayorov}, \citenamefont {Nelson}, \citenamefont {Jones}, \citenamefont
  {Kern}, \citenamefont {Larson}, \citenamefont {Carey}, \citenamefont {Polat},
  \citenamefont {Feng}, \citenamefont {Moore}, \citenamefont {{VanderPlas}},
  \citenamefont {Laxalde}, \citenamefont {Perktold}, \citenamefont {Cimrman},
  \citenamefont {Henriksen}, \citenamefont {Quintero}, \citenamefont {Harris},
  \citenamefont {Archibald}, \citenamefont {Ribeiro}, \citenamefont
  {Pedregosa}, \citenamefont {{van Mulbregt}},\ and\ \citenamefont {{SciPy 1.0
  Contributors}}}]{2020SciPy-NMeth}%
  \BibitemOpen
  \bibfield  {author} {\bibinfo {author} {\bibfnamefont {P.}~\bibnamefont
  {Virtanen}}, \bibinfo {author} {\bibfnamefont {R.}~\bibnamefont {Gommers}},
  \bibinfo {author} {\bibfnamefont {T.~E.}\ \bibnamefont {Oliphant}}, \bibinfo
  {author} {\bibfnamefont {M.}~\bibnamefont {Haberland}}, \bibinfo {author}
  {\bibfnamefont {T.}~\bibnamefont {Reddy}}, \bibinfo {author} {\bibfnamefont
  {D.}~\bibnamefont {Cournapeau}}, \bibinfo {author} {\bibfnamefont
  {E.}~\bibnamefont {Burovski}}, \bibinfo {author} {\bibfnamefont
  {P.}~\bibnamefont {Peterson}}, \bibinfo {author} {\bibfnamefont
  {W.}~\bibnamefont {Weckesser}}, \bibinfo {author} {\bibfnamefont
  {J.}~\bibnamefont {Bright}}, \bibinfo {author} {\bibfnamefont {S.~J.}\
  \bibnamefont {{van der Walt}}}, \bibinfo {author} {\bibfnamefont
  {M.}~\bibnamefont {Brett}}, \bibinfo {author} {\bibfnamefont
  {J.}~\bibnamefont {Wilson}}, \bibinfo {author} {\bibfnamefont {K.~J.}\
  \bibnamefont {Millman}}, \bibinfo {author} {\bibfnamefont {N.}~\bibnamefont
  {Mayorov}}, \bibinfo {author} {\bibfnamefont {A.~R.~J.}\ \bibnamefont
  {Nelson}}, \bibinfo {author} {\bibfnamefont {E.}~\bibnamefont {Jones}},
  \bibinfo {author} {\bibfnamefont {R.}~\bibnamefont {Kern}}, \bibinfo {author}
  {\bibfnamefont {E.}~\bibnamefont {Larson}}, \bibinfo {author} {\bibfnamefont
  {C.~J.}\ \bibnamefont {Carey}}, \bibinfo {author} {\bibfnamefont
  {{\.I}.}~\bibnamefont {Polat}}, \bibinfo {author} {\bibfnamefont
  {Y.}~\bibnamefont {Feng}}, \bibinfo {author} {\bibfnamefont {E.~W.}\
  \bibnamefont {Moore}}, \bibinfo {author} {\bibfnamefont {J.}~\bibnamefont
  {{VanderPlas}}}, \bibinfo {author} {\bibfnamefont {D.}~\bibnamefont
  {Laxalde}}, \bibinfo {author} {\bibfnamefont {J.}~\bibnamefont {Perktold}},
  \bibinfo {author} {\bibfnamefont {R.}~\bibnamefont {Cimrman}}, \bibinfo
  {author} {\bibfnamefont {I.}~\bibnamefont {Henriksen}}, \bibinfo {author}
  {\bibfnamefont {E.~A.}\ \bibnamefont {Quintero}}, \bibinfo {author}
  {\bibfnamefont {C.~R.}\ \bibnamefont {Harris}}, \bibinfo {author}
  {\bibfnamefont {A.~M.}\ \bibnamefont {Archibald}}, \bibinfo {author}
  {\bibfnamefont {A.~H.}\ \bibnamefont {Ribeiro}}, \bibinfo {author}
  {\bibfnamefont {F.}~\bibnamefont {Pedregosa}}, \bibinfo {author}
  {\bibfnamefont {P.}~\bibnamefont {{van Mulbregt}}}, and\ \bibinfo {author}
  {\bibnamefont {{SciPy 1.0 Contributors}}},\ }\href
  {https://doi.org/10.1038/s41592-019-0686-2} {\bibfield  {journal} {\bibinfo
  {journal} {Nature Methods}\ }\textbf {\bibinfo {volume} {17}},\ \bibinfo
  {pages} {261} (\bibinfo {year} {2020})}\BibitemShut {NoStop}%
\bibitem [{\citenamefont {{Macleod}}\ \emph {et~al.}(2021)\citenamefont
  {{Macleod}}, \citenamefont {{Areeda}}, \citenamefont {{Coughlin}},
  \citenamefont {{Massinger}},\ and\ \citenamefont {{Urban}}}]{gwpy}%
  \BibitemOpen
  \bibfield  {author} {\bibinfo {author} {\bibfnamefont {D.~M.}\ \bibnamefont
  {{Macleod}}}, \bibinfo {author} {\bibfnamefont {J.~S.}\ \bibnamefont
  {{Areeda}}}, \bibinfo {author} {\bibfnamefont {S.~B.}\ \bibnamefont
  {{Coughlin}}}, \bibinfo {author} {\bibfnamefont {T.~J.}\ \bibnamefont
  {{Massinger}}}, and\ \bibinfo {author} {\bibfnamefont {A.~L.}\ \bibnamefont
  {{Urban}}},\ }\href {https://doi.org/10.1016/j.softx.2021.100657} {\bibfield
  {journal} {\bibinfo  {journal} {SoftwareX}\ }\textbf {\bibinfo {volume}
  {13}},\ \bibinfo {pages} {100657} (\bibinfo {year} {2021})}\BibitemShut
  {NoStop}%
\bibitem [{\citenamefont {Hunter}(2007)}]{Hunter:2007}%
  \BibitemOpen
  \bibfield  {author} {\bibinfo {author} {\bibfnamefont {J.~D.}\ \bibnamefont
  {Hunter}},\ }\href {https://doi.org/10.1109/MCSE.2007.55} {\bibfield
  {journal} {\bibinfo  {journal} {Computing in Science \& Engineering}\
  }\textbf {\bibinfo {volume} {9}},\ \bibinfo {pages} {90} (\bibinfo {year}
  {2007})}\BibitemShut {NoStop}%
\bibitem [{\citenamefont {Waskom}(2021)}]{Waskom2021}%
  \BibitemOpen
  \bibfield  {author} {\bibinfo {author} {\bibfnamefont {M.~L.}\ \bibnamefont
  {Waskom}},\ }\href {https://doi.org/10.21105/joss.03021} {\bibfield
  {journal} {\bibinfo  {journal} {Journal of Open Source Software}\ }\textbf
  {\bibinfo {volume} {6}},\ \bibinfo {pages} {3021} (\bibinfo {year}
  {2021})}\BibitemShut {NoStop}%
\bibitem [{\citenamefont {Khan}\ \emph {et~al.}(2016)\citenamefont {Khan},
  \citenamefont {Husa}, \citenamefont {Hannam}, \citenamefont {Ohme},
  \citenamefont {P\"urrer}, \citenamefont {Jim\'enez~Forteza},\ and\
  \citenamefont {Boh\'e}}]{Khan:2015jqa}%
  \BibitemOpen
  \bibfield  {author} {\bibinfo {author} {\bibfnamefont {S.}~\bibnamefont
  {Khan}}, \bibinfo {author} {\bibfnamefont {S.}~\bibnamefont {Husa}}, \bibinfo
  {author} {\bibfnamefont {M.}~\bibnamefont {Hannam}}, \bibinfo {author}
  {\bibfnamefont {F.}~\bibnamefont {Ohme}}, \bibinfo {author} {\bibfnamefont
  {M.}~\bibnamefont {P\"urrer}}, \bibinfo {author} {\bibfnamefont
  {X.}~\bibnamefont {Jim\'enez~Forteza}}, and\ \bibinfo {author} {\bibfnamefont
  {A.}~\bibnamefont {Boh\'e}},\ }\href
  {https://doi.org/10.1103/PhysRevD.93.044007} {\bibfield  {journal} {\bibinfo
  {journal} {Phys. Rev. D}\ }\textbf {\bibinfo {volume} {93}},\ \bibinfo
  {pages} {044007} (\bibinfo {year} {2016})},\ \Eprint
  {https://arxiv.org/abs/1508.07253} {arXiv:1508.07253 [gr-qc]} \BibitemShut
  {NoStop}%
\bibitem [{\citenamefont {Finn}\ and\ \citenamefont
  {Chernoff}(1993)}]{Finn:1992xs}%
  \BibitemOpen
  \bibfield  {author} {\bibinfo {author} {\bibfnamefont {L.~S.}\ \bibnamefont
  {Finn}}and\ \bibinfo {author} {\bibfnamefont {D.~F.}\ \bibnamefont
  {Chernoff}},\ }\href {https://doi.org/10.1103/PhysRevD.47.2198} {\bibfield
  {journal} {\bibinfo  {journal} {Phys. Rev. D}\ }\textbf {\bibinfo {volume}
  {47}},\ \bibinfo {pages} {2198} (\bibinfo {year} {1993})},\ \Eprint
  {https://arxiv.org/abs/gr-qc/9301003} {arXiv:gr-qc/9301003} \BibitemShut
  {NoStop}%
\bibitem [{\citenamefont {O'Shaughnessy}\ \emph {et~al.}(2010)\citenamefont
  {O'Shaughnessy}, \citenamefont {Kalogera},\ and\ \citenamefont
  {Belczynski}}]{OShaughnessy:2009szr}%
  \BibitemOpen
  \bibfield  {author} {\bibinfo {author} {\bibfnamefont {R.}~\bibnamefont
  {O'Shaughnessy}}, \bibinfo {author} {\bibfnamefont {V.}~\bibnamefont
  {Kalogera}}, and\ \bibinfo {author} {\bibfnamefont {K.}~\bibnamefont
  {Belczynski}},\ }\href {https://doi.org/10.1088/0004-637X/716/1/615}
  {\bibfield  {journal} {\bibinfo  {journal} {Astrophys. J.}\ }\textbf
  {\bibinfo {volume} {716}},\ \bibinfo {pages} {615} (\bibinfo {year}
  {2010})},\ \Eprint {https://arxiv.org/abs/0908.3635} {arXiv:0908.3635
  [astro-ph.CO]} \BibitemShut {NoStop}%
\bibitem [{\citenamefont {Ade}\ \emph {et~al.}(2016)\citenamefont {Ade} \emph
  {et~al.}}]{Planck:2015fie}%
  \BibitemOpen
  \bibfield  {author} {\bibinfo {author} {\bibfnamefont {P.~A.~R.}\
  \bibnamefont {Ade}} \emph {et~al.} (\bibinfo {collaboration} {Planck}),\
  }\href {https://doi.org/10.1051/0004-6361/201525830} {\bibfield  {journal}
  {\bibinfo  {journal} {Astron. Astrophys.}\ }\textbf {\bibinfo {volume}
  {594}},\ \bibinfo {pages} {A13} (\bibinfo {year} {2016})},\ \Eprint
  {https://arxiv.org/abs/1502.01589} {arXiv:1502.01589 [astro-ph.CO]}
  \BibitemShut {NoStop}%
\bibitem [{\citenamefont {{Thrane}}\ and\ \citenamefont
  {{Talbot}}(2019)}]{Thrane_Talbot_2019}%
  \BibitemOpen
  \bibfield  {author} {\bibinfo {author} {\bibfnamefont {E.}~\bibnamefont
  {{Thrane}}}and\ \bibinfo {author} {\bibfnamefont {C.}~\bibnamefont
  {{Talbot}}},\ }\href {https://doi.org/10.1017/pasa.2019.2} {\bibfield
  {journal} {\bibinfo  {journal} {\pasa}\ }\textbf {\bibinfo {volume} {36}},\
  \bibinfo {eid} {e010} (\bibinfo {year} {2019})},\ \Eprint
  {https://arxiv.org/abs/1809.02293} {arXiv:1809.02293 [astro-ph.IM]}
  \BibitemShut {NoStop}%
\end{thebibliography}%
\bibliographystyle{apsrev4-2}

\end{document}